\documentclass[twocolumn]{aastex62}
\usepackage{float,graphicx,amsmath,multirow,mathtools,ulem,comment}
\usepackage{color}
\usepackage[caption=false]{subfig}
\newcommand*\chem[1]{\ensuremath{\mathrm{#1}}}
\newcommand{\kms}{km~s$^{-1}$ }
\newcommand{\cmq}{cm$^{-3}$ }
\newcommand{\cms}{cm$^{-2}$}

\received{Oct 2018}
\revised{Jan 2019}
\accepted{\today}
\submitjournal{ApJ}

\shorttitle{Methyl carbamate toward hot corinos}
\shortauthors{Sahu et al.}

\begin{document}

\title{Constraints of the formation and abundances of methyl carbamate, a glycine isomer, in hot corinos}

\author{Dipen sahu}
\affiliation{Academia Sinica Institute of Astronomy and Astrophysics, 11F of AS/NTU Astronomy-Mathematics Building, No.1, Sec. 4, Roosevelt Rd, Taipei 10617, Taiwan, R.O.C.}
\author{Sheng-Yuan Liu}
\affiliation{Academia Sinica Institute of Astronomy and Astrophysics, 11F of AS/NTU Astronomy-Mathematics Building, No.1, Sec. 4, Roosevelt Rd, Taipei 10617, Taiwan, R.O.C.}
\author{Ankan Das}
\affiliation{Indian Centre for Space Physics, 43 Chalantika, Garia St. Road, Kolkata 700084, India}
\author{Prasanta Garai}
\affiliation{Indian Centre for Space Physics, 43 Chalantika, Garia St. Road, Kolkata 700084, India}
\author{Valentine Wakelam}
\affiliation{Laboratoire d'astrophysique de Bordeaux, Univ. Bordeaux, CNRS, B18N, all\'ee Geoffroy Saint-Hilaire, 33615 Pessac, France}
\correspondingauthor{Dipen Sahu}
\email{dipenthink@gmail.com}

\begin{abstract}
 Methyl carbamate CH$_3$OC(O)NH$_2$ is an isomer of glycine. 
 Quantum chemical analyses show that methyl carbamate  is more stable isomer than glycine. 
 Because of this,  there could be a higher chance for methyl carbamte to exist in the interstellar medium as compared to glycine. 
 Despite immense searches, till now glycine  has not been detected in the ISM, therefore it is worthwhile to search its isomer methyl carbamate. 
 In this paper, we present the constraints of methyl carbamate formation under the interstellar conditions.
 Large complex organic molecules are favorably produced in hot-corino environments of low mass protostars. 
 We  for the first time carried out astrochemical modeling focusing on the formation of methyl carbamate  in physical conditions similar to hot-corino objects. 
 Consequently, we examined ALMA archival data for existing spectral line observations toward hot corinos NGC1333 IRAS 4A2 and IRAS 16293B. 
 Within the common spectral range towards these sources, we found three features   are possibly related to the  spectral transitions of  methyl carbamate  and consequently  estimate the  upper limit of  column densities. 
 Results of chemical modeling are consistent with the observational upper limit of estimated column density/abundance toward the sources.
  This may hint the validation of the proposed formation mechanism.
 Future observations using telescope like ngVLA may confirm the presence of MC toward the hot corinos.

\end{abstract}


\noindent

\section{Introduction} \label{sec:intro}
Amino acids are the building blocks of life as it is the essential compound for protein formation.
\citet{miller53} from his experiments showed that many amino acids can be produced in the laboratory using electric discharge in an aqueous mixture of ammonia, methane, and dihydrogen.
In a later examination, \citet{khare86} observed 16 amino acids from the acidic treatment of tholins, which is present in Titan's atmosphere.
Various investigations have also been performed to detect amino acids in extraterrestrial objects. 
For example, a number of amino acids have been detected in the Murchison meteorite \citep{toshiki17}.
Amino acids have also been found in comet 67P/Churyumov-Gerasimenko \citep{altwegg16}. 
All these studies suggest that the presence of amino acids could be quite ubiquitous in extraterrestrial environments.

To further explore the prebiotic chemistry, it is indispensable to study the simplest amino acids, glycine. Although extensive observational searches had been made, no conclusive detection of glycine in the interstellar medium (ISM) is reported \citep[e.g.,][]{kuan03, snyder05}.
Various authors discussed the formation of glycine in the ISM using possible pathways and chemical modeling \citep[e.g.][]{garrod13,suzuki18}, but it remains illusive.

In the context of isomeric species, it is interesting to note that ISM molecules often form preferentially following the `minimum energy principle (MEP)' \citep{guillemin14}.
The MEP predicts that thermodynamically the most stable isomer should be the most abundant. 
As an example, in the molecular clouds where \chem{C_2H_3N} isomers have been detected - the more stable acetonitrile (\chem{CH_3CN}) is found to be more abundant than  methyl isocyanide(\chem{CH_3NC}) and ketenimine (\chem{CH_2CNH}).
Similarly acetaldehyde (\chem{CH_3CHO}) is more abundant than its isomeric counterpart  vinyl alcohol (\chem{CH_2CHOH}) \citep[see][and references therein]{guillemin14}.
There could exist, however, some exceptions due to the difference in the efficiencies of their reactions \citep{lattelais09}. 
 \citet{loomis2015} further argued that MEP may not be generally applicable to interstellar chemistry and the formation of complex organic molecules(COMs). 
They suggested that in the ISM local environments play a major role for the formation of chemical species rather than thermodynamic stability of individual species.
Neverthless, it is worthwhile to examine the molecule of prebiotic important \chem{C_2H_5O_2N}, of which three isomers are glycine (\chem{NH_2CH_2COOH}),  N-methylcarbamic acid (\chem{CH_3NHCOOH}) and methyl carbamate (\chem{CH_3OC(NH_2)O}).
Quantum chemical studies of the energy stability showed that N-methylcarbamic acid (dipole moment $\sim 1.2$ Debye ) is the most stable isomer among these three \citep{lattelais11}.
To the best of our knowledge,  N-methylcarbamic acid has never been searched in the ISM, maybe due to the unavailability of its  spectroscopic data.
{On the other hand, methyl carbamate (MC)} is the second stable isomer, $~4$ kcal/mole above the most stable isomer (\chem{CH_3NHCOOH}) while glycine is $\sim 10$ kcal/mol above \citep{lattelais11}.
 In addition,  MC  \footnote{Polarizability $\rm{6.6 \pm 0.5 \times 10^{24}cm^3}$, \url{CSID:11229, http://www.chemspider.com/Chemical-Structure.11229.html (accessed 02:52, Jun 24, 2020)}} has a dipole moment of $\sim 2.4$ Debye, higher than N-methylcarbamic.
Maybe for this reason, MC was searched toward the intermediate-mass protostar IRAS21391+58502 and the hot core W51e2 \citep{demyk04}.  However, no further detail or positive results are available in the literature.

 Nonetheless, in the ISM, COMs formation are often described to be formed predominantly on the ice phase via radical-radical reactions \citep[e.g.,][]{garrod13}. Motivated by these,
 we performed chemical modeling assuming that MC is formed through similar pathways and searched
for possible signatures of MC in the protostellar hot corino sources NGC~1333 IRAS~4A2 and IRAS 16293-2422 B.
Hot corinos are associated with low-mass star formation at its relatively early evolutionary stage.
Due to the elevated temperature ( $>$ 100~K) in the vicinity of protostars, molecules in icy mantels come off to the gas phase due to thermal desorption.
For this reason, a large abundance of COMs are often observed in the gas phase  toward hot-corino objects.
Results from chemical modeling show that COMs formation more likely occurs during the warm-up phase in the star formation process \citep{garrod06}.
In this context, the hot-corino sources are good target to search for COMs such as MC.
We model the formation of methyl carbamate in the hot-corino environments to  study the possibility of its presence. 
We also present the ALMA observations of NGC~1333 IRAS~4A2 and IRAS 16293-2422 B  towards of which we derived MC's column density upper limits being consistent with the chemical modeling results.

\section{Astrochemical modeling} \label{sec:model}
\subsection{The chemical model, NAUTILUS}
To estimate the formation of methyl carbamate in the ISM and particularly in the hot corino environment,
we use the NAUTILUS chemical code.
The details of its formalism are described in \citet{ruaud16}.
Using  NAUTILUS one can compute the time-evolution of chemistry (called 0D) and also spatial distributions of species \citep[called 1D; see][ for example]{wakelam14}.
The  chemical code considers the interaction between gas and grain and calculates the chemical evolution both in gas and on dust surfaces.
The rate equation approximation is used in the code to 
 estimate molecular abundances of species  
 in the gas phase, dust surfaces and in the bulk of grain mantle (i.e., a three-phase model).
For the dust surface chemistry, NAUTILUS considers only the physisorption of chemical species onto the dust surface and subsequent diffusion mechanism to produce new species due to chemical reactions/recombinations.
The surface species can come to the gas phase via different desorption mechanisms  including thermal desorption (temperature dependent), UV photodesorption, cosmic-ray induced evaporation, and chemical desorption \citep[see] [and references therein]{ruaud16}.

 \begin{table}[]
     \centering
     \caption{  Initial abundances with respect to total hydrogen nuclei}
     \begin{tabular}{|l|l|}
     \hline
      Species    & Abundance  \\
      \hline
      H$_2$  & $5 \times 10^{-1}$ \\
     He & $1.0 \times 10^{-1}$\\
     O  & $1.76 \times 10^{-4}$ \\
     \chem{C^+} & $ 7.3 \times 10^{-5}$\\ 
     N  & $ 2.14 \times10^{-5} $\\
    \chem{S^+} & $8 \times 10^{-8}$ \\
  \chem{Si^+} & $8 \times 10^{-9}$ \\
   \chem{Mg^+} & $7 \times 10^{-9}$ \\
    \chem{Cl^+} & $4 \times 10^{-9}$ \\
     \chem{Fe^+} & $3 \times 10^{-9}$ \\
   \chem{P^+} & $3 \times 10^{-9}$ \\
  \chem{Na^+} & $2 \times 10^{-9}$ \\
\hline
     \end{tabular}
     
     \label{tab:element}
 \end{table}

\subsection{chemical network and methyl carbamate} \label{subsec:network}
 We developed a chemical network for the methyl carbamate starting from the kida.uva.2014 gas-phase network \footnote{\url{(http://kida.astrophy.u-bordeaux.fr)}} and from the surface network described in \citet{ruaud16} for the surface processes, both updated from \citet{wakelam15,wakelam17,hickson16,loison16,loison17,vidal2017}. The starting network is the same network used in \citet{wakelam2019}.
To these sets of network, we have added several relevant species and reactions for exploring the formation of MC. 
The added species are \chem{s-CH_3O}, \chem{s-NH_2CO}, \chem{s-H_2NCO_2CH_3} (MC), \chem{~s-NH_2},              \chem{~s-CH_3OCO}, 
\chem{~s-CH_3CO}, and \chem{ ~s-HOCO} in the grain phase, with `s' representing surface species. 
Along with these, gas-phase species corresponding to the grain phase species including
some newly produced gas phase species are also considered.  Initial abundances of elemental species \citep{das15a} and reactions involving these species are summarized in  Table ~\ref{tab:element} and Table ~\ref{tab:network}, respectively.

\begin{table*}[ht]
\centering
\scriptsize
\caption{  The chemical reaction network of methyl carbamate} 
\begin{tabular}{l|c|c|c|c}
\hline
     
\hline
Reactions & $\alpha$ & $\beta$ & $\gamma$ & Reaction type\\
\hline
 \chem{s-CH_3O} + \chem{s-NH_2CO} $\rightarrow$ \chem{s-H_2NCO_2CH_3} & 1.000e+00 & 0.000e+00 & 0.000e+00 & surface  \\
 \chem{s-H_2NCO_2CH_3}  $\rightarrow$ \chem{H_2NCO_2CH_3} & 1.000e+00 &  0.000e+00 & 0.000e+00   & surface\\
 \hline 
\chem{s-NH_2} + \chem{s-CO} $\rightarrow$ \chem{s-NH_2CO} & 1.000e+00 & 0.000e+00 & 2.100E+03 & surface \\
\chem{s-NH_2CO} + \chem{s-H} $\rightarrow$ \chem{s-NH_2CHO}& 1.000e+00 & 0.000e+00 & 0.000e+00 & surface \\
\chem{s-NH_2CO} + \chem{s-H} $\rightarrow$ \chem{s-HNCO} + \chem{H_2}& 1.000e+00 & 0.000e+00 & 0.000e+00 & surface \\
\chem{s-NH_2CO} + \chem{s-H} $\rightarrow$ \chem{s-NH_3} + \chem{CO}& 1.000e+00 & 0.000e+00 & 0.000e+00 & surface \\
\chem{s-NH_2CHO} + \chem{s-H} $\rightarrow$ \chem{s-NH_2CO} + \chem{s-H_2}& 1.000e+00 & 0.000e+00 & 2.500e+03 & surface \\
\chem{s-HNCO} + \chem{s-H} $\rightarrow$ \chem{s-NH_2CO} & 1.000e+00 & 0.000e+00 & 4.150e+03 & surface \\

\chem{s-NH_2CO} $\rightarrow$ \chem{NH_2CO}& 1.000e+00 & 0.000e+00 & 0.000e+00 & surface\\
 \hline
 \chem{H_2NCO_2CH_3} +  Photon     $\rightarrow$ \chem{NH_2CO} +\chem{CH_3O} & 1.380e-09 & 0.000e+00 & 1.730e+00 & photo \\
\chem{H_2NCO_2CH_3} +  CRP   $\rightarrow$  \chem{NH_2CO} + \chem{CH_3O} &    1.500e+03 & 0.000e+00 & 0.000e+00  & CRP\\
\chem{H_2NCO_2CH_3} +  Photon  $\rightarrow$ \chem{NH_2} + \chem{CH_3OCO} & 1.380e-09 & 0.000e+00 & 1.730e+00  & photo\\
\chem{H_2NCO_2CH_3} +  CRP   $\rightarrow$  \chem{NH_2} + \chem{CH_3OCO} &   1.500e+03 &  0.000e+00 & 0.000e+00 & CRP\\
\chem{H_2NCO_2CH_3} +  \chem{C^+}  $\rightarrow$   C + \chem{COOCH_4^+} +    NH & 1.860e-09 &   1.000e+00 &  3.240e+00 & Bi-mol \\
\chem{H_2NCO_2CH_3} + \chem{He^+} $\rightarrow$  He + \chem{CH_3} +\chem{H_2NCO_2^+} &  3.080e-09 &  1.000e+00 &   3.240e+00 & Bi-mol\\
\chem{H_2NCO_2CH_3} + \chem{H_3^+} $\rightarrow$  \chem{H_2} + \chem{NH_6C_2O_2^+}  &  3.510e-09 &  1.000e+00 &   3.240e+00 &Bi-mol \\
\chem{H_2NCO_2CH_3} +  \chem{H^+}  $\rightarrow$    \chem{NH_2} +  \chem{COOCH_4^+} &  6.010e-09 &  1.000e+00 &    3.240e+00 & Bi-mol\\
\chem{H_2NCO_2CH_3} +\chem{HCO^+}   $\rightarrow$   CO + \chem{NH_6C_2O_2^+}  &  1.310e-09 &  1.000e+00 &   3.240e+00 & Bi-mol\\
\chem{H_2NCO_2CH_3} +\chem{H_3O^+} $\rightarrow$  \chem{H_2O} +\chem{NH_6C_2O_2^+}&  1.530e-09 &  1.000e+00 &   3.240e+00 & Bi-mol\\
 \hline
 \chem{CH_3} + HOCO $\rightarrow$  \chem{CH_3OCO} +   H &  1.000e-10 & 0.000e+00 &  8.004e+03 & Bi-mol\\
\chem{H_3^+} + \chem{CH_3OCO} $\rightarrow$  \chem{CH_3COOH^+} +\chem{H_2}&     1.000e-09 & -5.000e-01 &  0.000e+00 & Bi-mol \\
\chem{HCO^+} + \chem{CH_3OCO}   $\rightarrow$  \chem{CH_3COOH^+} + CO &       1.030e-09 & -5.000e-01 & 0.000e+00 & Bi-mol\\
\chem{H^+} + \chem{CH_3OCO} $\rightarrow$ \chem{CH_3OCO^+} +  H  &             1.000e-09 & -5.000e-01 & 0.000e+00 & Bi-mol\\
\chem{CO^+} +  \chem{CH_3OCO}  $\rightarrow$  \chem{CH_3OCO^+} + CO &           1.440e-09 & -5.000e-01 &  0.000e+00 & Bi-mol\\
\chem{He^+} +  \chem{CH_3OCO} $\rightarrow$ \chem{CH_3OCO^+} + He             & 1.000e-09 & -5.000e-01 &  0.000e+00 &Bi-mol\\
\chem{CH_3COOH^+} +  e$^-$  $\rightarrow$ \chem{CH_3OCO} + H                   & 3.000e-07 & -5.000e-01 & 0.000e+00 &Bi-mol\\
\chem{CH_3OCO}  + CRP $\rightarrow$ \chem{CO_2} +  \chem{CH_3} &               4.000e+03 & 0.000e+00 & 0.000e+00 & CRP\\
\chem{CH_3OCO} + Photon $\rightarrow$  \chem{CO_2} + \chem{CH_3} &              5.000e-10 &  0.000e+00 & 0.000e+00 & photo\\

\chem{CH_3OCO^+} +  e$^-$ $\rightarrow$ \chem{CH_3} +\chem{CO_2}               & 1.500e-07 & -5.000e-01 & 0.000e+00 &Bi-mol\\
\chem{H_2NCO_2^+} +  e$^-$ $\rightarrow$  \chem{NH_2} + \chem{CO_2} &         1.500e-07 & -5.000e-01 &  0.000e+00 & Bi-mol\\
 \hline
\chem{NH_6C_2O_2^+} +  e$^-$ $\rightarrow$  H  + \chem{H_2NCO_2CH_3} &          1.500e-07 & -5.000e-01 & 0.000e+00 & Bi-mol\\
\chem{NH_6C_2O_2^+} +  e$^-$  $\rightarrow$ \chem{NH_2CO} + \chem{CH_3OH}     & 1.500e-07 & -5.000e-01 & 0.000e+00 & Bi-mol\\
\hline
\chem{NH_2CHO} +   H $\rightarrow$ \chem{H_2} + \chem{NH_2CO}  & 1.000e-10 & 0.000e+00 &  3.1e+3/2.4e+3 ** & Bi-mol \\
\chem{NH_2CO} +   H $\rightarrow$ \chem{H_2} + \chem{HNCO}  & 1.000e-10 & 0.000e+00 & 0.000e+00 & Bi-mol\\
\chem{HNCO} +   H $\rightarrow$ \chem{NH_2CO}  & 1.000e-10 & 0.000e+00 & 5.050e+03 & Bi-mol\\
\chem{NH_2CO} +   H $\rightarrow$ \chem{NH_2CHO}  & 1.000e-10 & 0.000e+00 & 0.000e+00 & Bi-mol\\

\chem{NH_2CO} +   CRP $\rightarrow$ \chem{NH_2} + CO  & 9.500e+03 & 0.000e+00 &  0.000e+00 & CRP\\
\chem{NH_2CO} + \chem{He^+} $\rightarrow$   He  + CO  + \chem{NH_2^+}  & 5.000e-01 &  2.450e-09 &  8.130e+00 & Bi-mol*\\
\chem{NH_2CO} +\chem{H_3^+} $\rightarrow$  \chem{H_2} + \chem{NH_2COH^+} & 1.000e+00  & 2.790e-09 & 8.130e+00 &Bi-mol* \\
\chem{NH_2CO} + \chem{HCO^+} $\rightarrow$   CO + \chem{NH_2COH^+} &  1.000e+00 &  1.130e-09 & 8.130e+00 &Bi-mol* \\
\chem{NH_2CO} + \chem{H^+}  $\rightarrow$  \chem{NH_2} +  \chem{HCO^+} & 5.000e-01  & 4.730e-09 &  8.130e+00 &Bi-mol* \\
\chem{NH_2CO} + \chem{H^+}  $\rightarrow$  HCO +\chem{NH_2^+} & 5.000e-01 & 4.730e-09 &  8.130e+00 &Bi-mol* \\ 
\chem{NH_2COH^+} + \chem{e^-}  $\rightarrow$    HCO + \chem{NH_2} &  1.500e-07 & -5.000e-01 &  0.000e+00 & Bi-mol\\
\chem{NH_2COH^+} + \chem{e^-}   $\rightarrow$   H + \chem{NH_2CO} & 1.500e-07 & -5.000e-01  & 0.000e+00 & Bi-mol\\ 
\chem{NH_2CHO}  +  H$\rightarrow$ \chem{H_2} +\chem{NH_2CO}  & 1.000e-10 &  0.000e+00  & 1.500e+03 &  Bi-mol* \\
\chem{NH_2CO}  +  H$\rightarrow$ \chem{H_2} +\chem{HNCO}  & 1.000e-10 &  0.000e+00  & 0.000e+00 &  Bi-mol* \\
\chem{NH_2CO}  +  H$\rightarrow$  \chem{NH_2CHO}  & 1.000e-10 &  0.000e+00  & 0.000e+00 &  Bi-mol* \\
\chem{HNCO}  +  H$\rightarrow$  \chem{NH_2CO}  & 1.000e-10 &  0.000e+00  & 3.000e+03 &  Bi-mol* \\
\hline
\multicolumn{5}{c}{Gas phase reaction
parameterized}\\
\hline
\chem{CH3}  +  HNCO $\rightarrow$  \chem{CH_3NCO} + H  & 1.000e-11 & 0.000e+00 & 0.000e+00 & Bi-mol  \\
\chem{NH_2} +   \chem{H_2CO} $\rightarrow$ \chem{NH_2CHO} + H  & 1.000e-12 & -2.560e+00 & 4.880e+00 & Bi-mol \\
\hline
\multicolumn{5}{p{16cm}}{Note: $\alpha$, $\beta$, $\gamma$ are rate constants, and rate coefficients are calculated based on reaction types as mentioned in column 5. The formulae that are used to calculate rate coefficients different categories of reactions are the following, CRP: $\kappa= \alpha \zeta$, where $\zeta$ is cosmic-ray ionization rate; photo: $\kappa = \alpha exp(-\gamma A_v)$;  Bi-mol: $\kappa(T)= \alpha (T/300)^\beta exp(-\gamma T)$; and Bi-mol*: $\kappa(T)= \alpha \beta(0.62 + 0.4767 \gamma (300/T)^{0.5})$. **We consider this values for model 1 and model 2 respectively. For `surface' reactions, the diffusion barrier to binding energy ratio is considered to be  0.5. } 
\end{tabular}
\label{tab:network}
\end{table*}

Carbamate compound is formally derived from carbamic acid (\chem{NH_2COOH}). In terrestrial conditions, methyl carbamate is produced by the reaction of methanol and urea.
This reaction, however may not be suitable for interstellar  medium due to its extreme conditions such as very low pressure and temperature compared to the Earth atmosphere. 
For efficient  formation of MC under ISM conditions, we consider the solid phase radical-radical reaction of \chem{CH_3O} and \chem{NH_2CO}:
\begin{equation}
    \chem{CH_3O} + \chem{NH_2CO} \rightarrow \chem{H_2NCO_2CH_3}
\end{equation}
We check the potential of this radical-radical reaction using quantum chemical calculations (B3LYP/6-311g++(d,p) method, Gaussian 09 software).
The enthalpy of the reaction is found to be $-80.79$ Kcal/mol, i.e. the reaction is exothermic and can proceed on grain surface, Additionally, we assume the radical-radical reaction to be barrier-less. 
The methoxy radical \chem{CH_3O} has been observed in the ISM \citep{chernicharo12} and \chem{NH_2CO} can form quite easily on grain surfaces \citep{ligter18}.  Pathways for \chem{CH_3O} are already available in the network, and pathways for \chem{NH_2CO} are shown in Table~\ref{tab:network}. \chem{CH_3O} mainly forms on grain surfaces via recombination reactions \citep[simplest molecule H$_2$ as well as COMs can form via this mechanism, see][]{sahu15,das16}  of atoms (e.g., O) and radicals (e.g., \chem{CH_3});  also it can be produced from  \chem{CH_3OH} as a result of hydrogen abstraction on grain surfaces.
\chem{CH_3O} is already present in our network and available in KIDA. Also, \chem{NH_2CO} mainly form on grain surfaces via recombination reactions  \citep{quenard2018}  and hydrogen abstraction from species like \chem{NH_2CHO}  \citep{haupa2019}.
The binding energies of \chem{NH_2CO} and \chem{CH_3O} are considered to be 5160 K and 4400 K, respectively \citep{wakelam17}.
Due to the high surface binding energies of the reactants, the radical-radical reaction (1) is unattainable at low temperature. 
At warm temperature ($\sim 100$~K and higher) the radicals get enough mobility to activate the reaction on grain surfaces.

Another tentative reaction that can produce MC in the solid phase is: 
\begin{equation}
    \chem{CH_3OCO} + \chem{NH_2} \rightarrow \chem{H_2NCO_2CH_3}
\end{equation}

This reaction was discussed by \citet{lee09}, where they speculated \chem{CH_3OCO} formation from the recombination of \chem{CH_3O} and CO.
They concluded that the reaction is unlikely  under cold condition. 
Though this reaction may be relevant for hot-corino like temperature ($>$ 100 K), we did not include this reaction in our network as the formation of \chem{CH_3OCO} radical is not well known.
Therefore, we have only considered  Reaction (1)  for solid phase production of MC. 
Due to  various surface desorptions the solid phase MC can populate the gas phase, while no other gas phase production of MC is considered.  We consider
different kinds of desorption mechanisms (see section 2.1) and find that thermal desorption is the dominant mechanism for gas phase abundance of MC. As shown in Table~\ref{tab:network}, the destruction of MC occurs mainly due to ion-neutral reactions and photo-dissociation including cosmic ray dissociation. We have also considered MC being adsorbed back onto grain surface in our network.

 From Table~\ref{tab:network}, we can see that \chem{NH_2CO} formation is related to \chem{NH_2CHO} (formamide) and HNCO (isocyanic acid). 
Also,  \chem{CH_3NCO} (methyl isocyanate) is another COM which is chemically related to HNCO \citep[e.g.,][]{gorai20}.
As MC mainly from via \chem{NH_2CO} radical, for the completeness of the chemical network we have to consider reactions related to \chem{NH_2CHO}, HNCO, and \chem{CH_3NCO}. 
Reaction pathways of these molecules are included in our network. 
In Table~2, we only noted some major reactions related to these molecules. 
Values of reactions rate constants vary over published articles. 
In our chemical network, we parameterized the major reactions (Table~\ref{tab:network}) following \citet{gorai20} and references therein. In Table~\ref{tab:network}, the gas phase reaction rate constants for MC are assumed to be similar to those for \chem{HCOOCH_3}. 
We consider only typical ion-neutral gas phase destruction pathways for MC. 
Also, tentative pathways for the newly produced species (e.g., \chem{CH_3OCO}, \chem{CH_3COOH^+}) are included for the completeness of the chemical network. 
Reaction rate constants are assumed similar to \chem{ CH_3NCO} for \chem{CH_3OCO} and related species.
We have added a few gas-phase destruction pathways (ion-neutral) for \chem{NH_2CO} following \chem{NH_2CN}. Gas phase reactions (neutral-neutral) of \chem{NH_2CO} and H are taken from \citet{gorai20}. 

 Since the reaction pathways for the main constituents of MC (\chem{NH_2CO} and \chem{CH_3O}) are incorporated, the network can be considered complete for investigating the MC formation.
If there are other reaction routes for the production of MC, in that case, assuming no major destruction of MC from unknown pathways, the current network would provide at least an estimate of the lower limit of MC abundance using the astrochemical modeling.

\subsection{Physical models}
To simulate the chemistry of MC, we consider the inner region of a low mass star forming core from its collapsing parent molecular cloud core to the subsequent warm-up phase due to YSO heating. 
Two physical models are adopted for mimicking the emergence of a hot corino.
Model~1 is a simple toy model in which the  cloud core is initially assumed to be at 8 K  with a density of $2\times10^4$ \cmq, and it evolves for $10^6$  years in the prestellar phase.
During the collapse phase, the density increases linearly to $10^8$~\cmq over a timescale of  $\sim 2.1 \times10^5$ years.
At the end of the collapse phase, the  dust temperature rises to 300 K (warm-up phase) within this period \citep[e.g.][]{viti04,garrod06}. The temperature during the warm-up phase is assumed to increase linearly with time. At this phase, dust and gas is well coupled, and we assume  gas and dust temperatures to be the same. 

We further draw, as Model~2, the physical model from \citet[][Fig 5 for 15 AU, in their work]{aikawa08}
which is based on a somewhat more realistic one-dimensional numerical radiation-hydrodynamical calculation. In Model 2, a static density ($2.0\times10^4$ \cmq at 8 K) evolution for $10^6$ years is assumed before the collapse and warm-up of the parent molecular cloud.  The warm-up time in this model is very short ($\sim 9\times10^4$  years ) compared to Model 1. Additionally, it warms up very quickly,  though the final temperature is $\sim 253$~K, it reaches $\sim 228$~K only after a warm-up of $2.0\times10^4$ years. Dust and gas temperature are considered to be the same throughout the evolution.

Depending on the gas density in the molecular cloud the extinction coefficient (\chem{A_V}) may vary, here we consider \chem{A_V} varies according to the relation \citep[][and references therein]{das15}:
\begin{equation}
   \chem{ n_H= n_{H0}[1 +(\sqrt{n_{Hmax}/n_{H0}} -1)A_V/A_{Vmax}]^2}
\end{equation}
where,  \chem{n_{H0}} is the initial density and  
\chem{n_{Hmax}} is the maximum density of  the molecular cloud;   \chem{A_{Vmax}} is the maximum visual co-efficient deep inside the cloud corresponding to the maximum density. In both our physical models, \chem{A_{V}}  is assumed to reach a maximum value 150 from a initial value of 10. 
A constant cosmic-ray ionization rate ($\zeta$) of $1.3 \times 10^{-17}$ s$^{-1}$ is assumed throughout the simulations.

\subsection{Results of modeling}

\begin{figure*}
    \subfloat[ Model 1\label{}]{%
       \includegraphics[width=9cm]{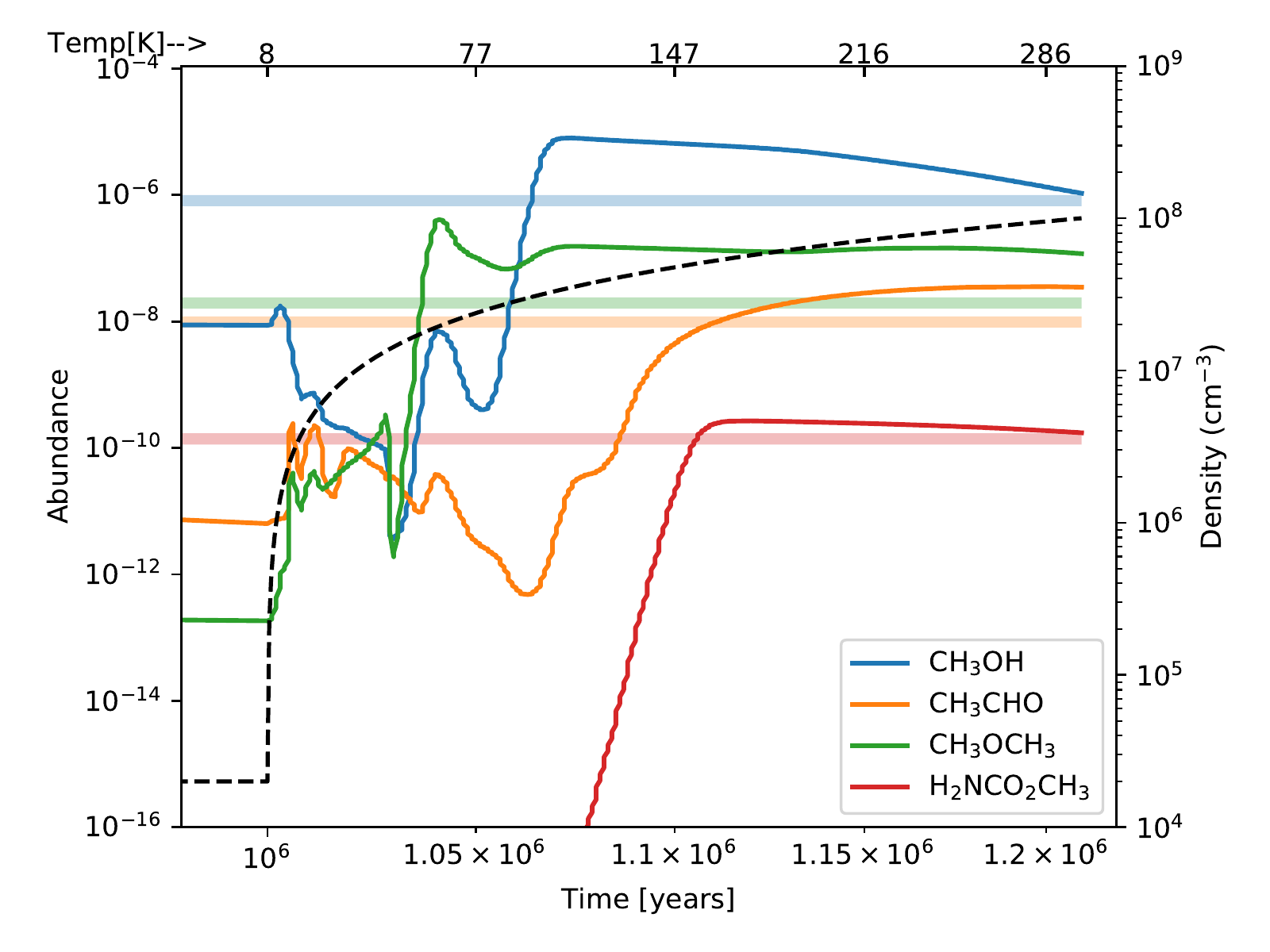}}
       \hfill
    \subfloat[ Model 1\label{}]{%
       \includegraphics[width=9cm]{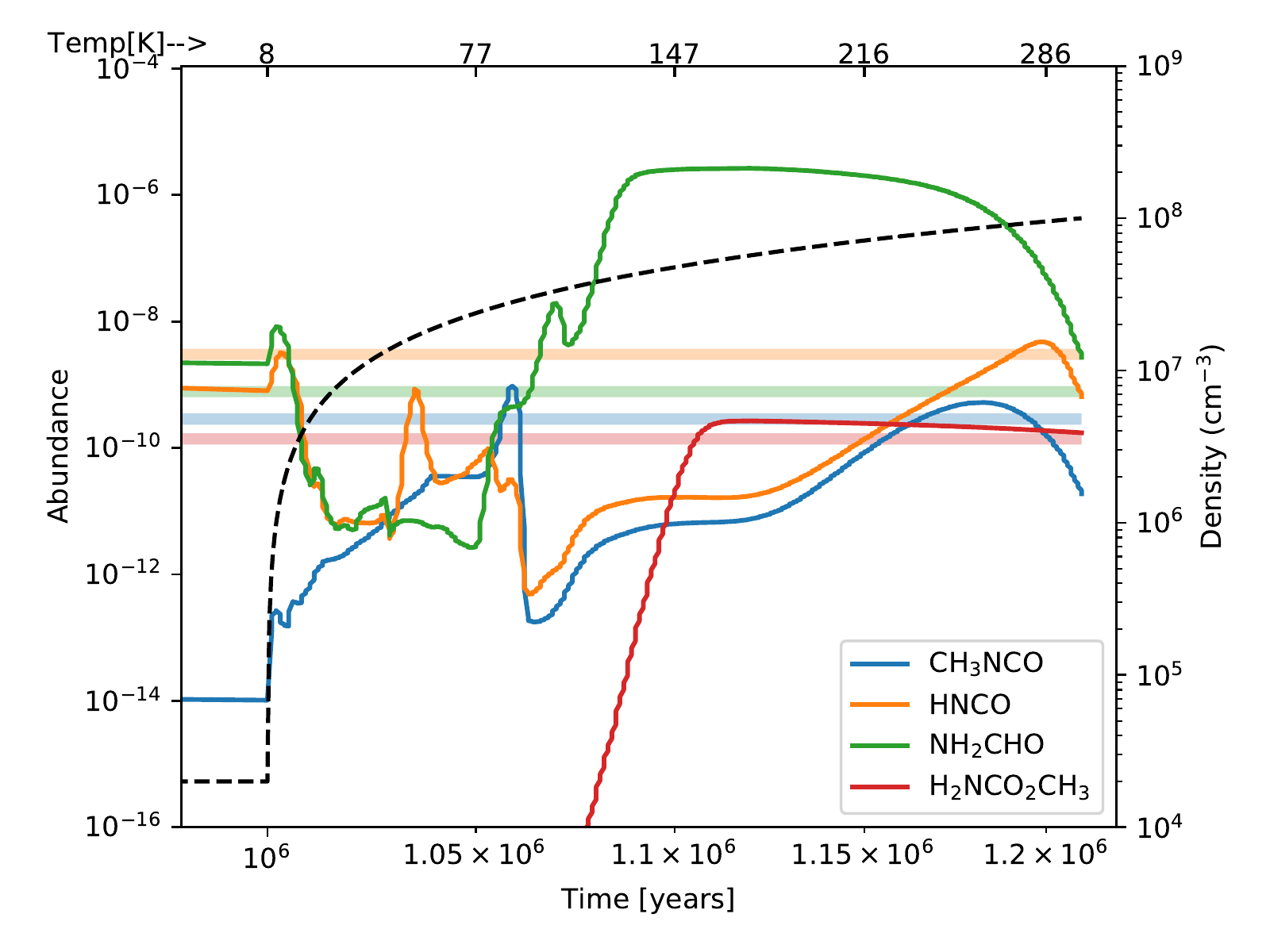}}
       \hfill
    \subfloat[ Model 2\label{}]{%
       \includegraphics[width=9cm]{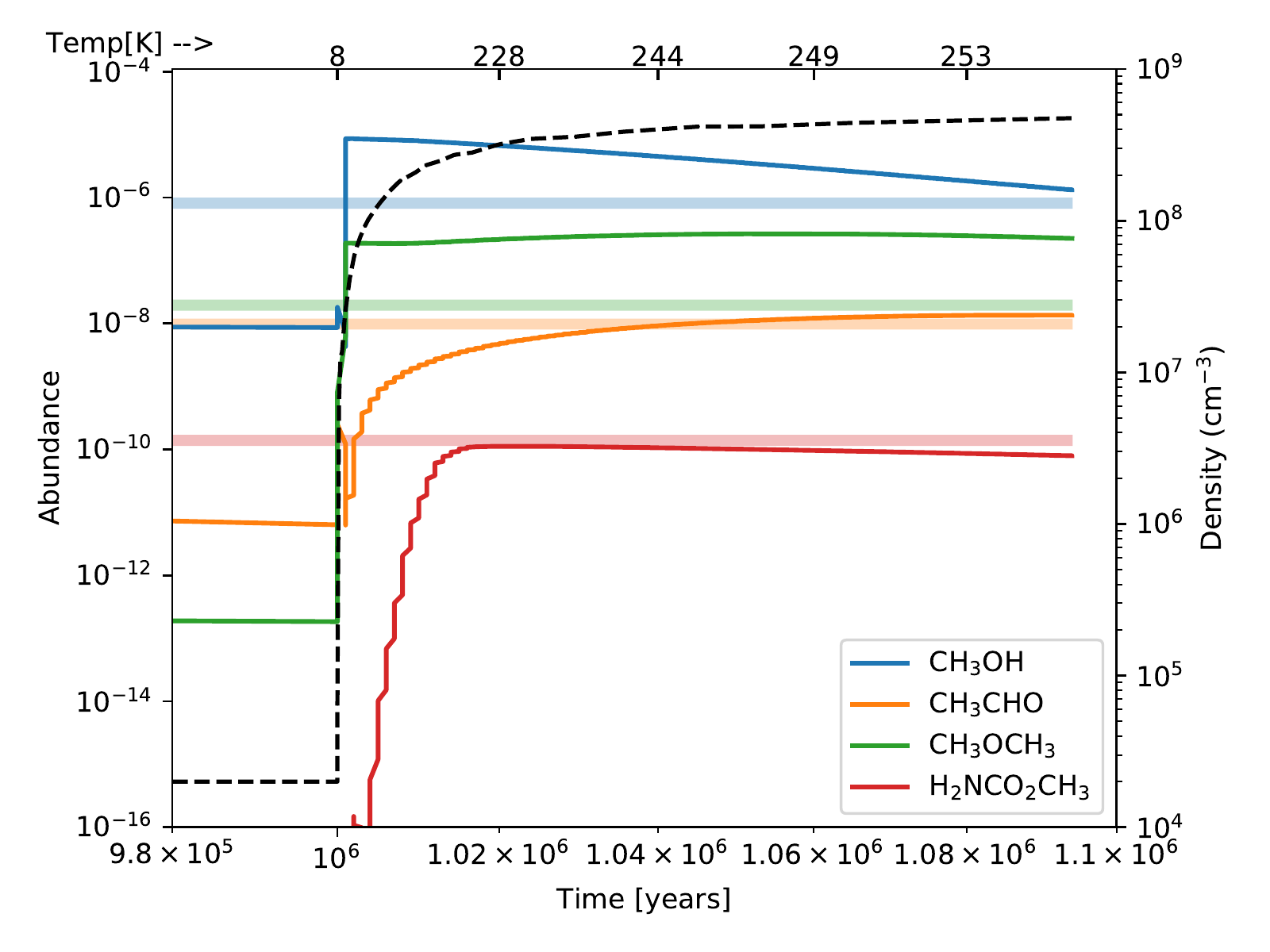}}
       \hfill
    \subfloat[ Model 2\label{}]{%
       \includegraphics[width=9cm]{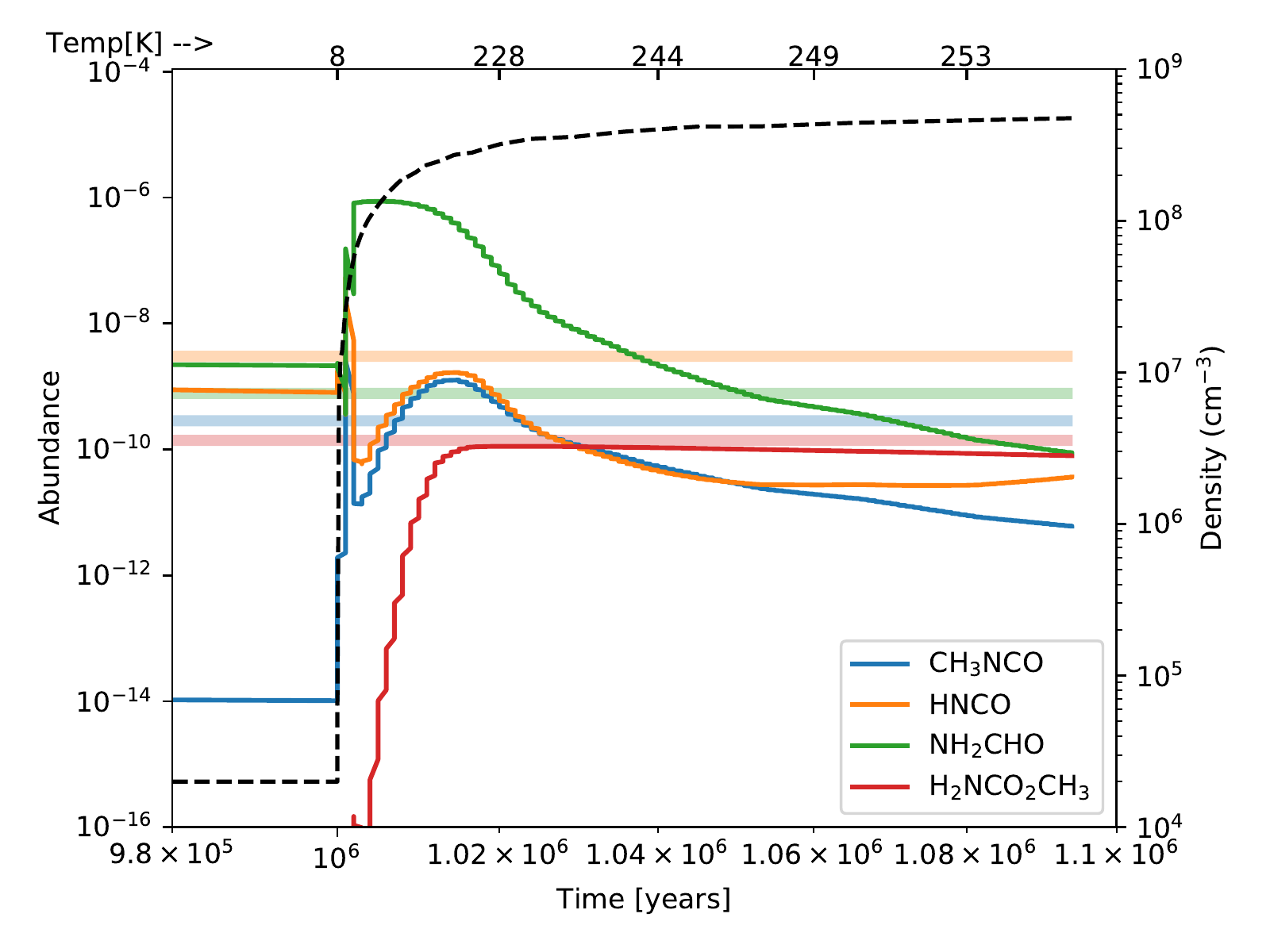}}
\caption{ Time evolution of chemical species during the warm-up phase (after an initial period of $10^6$ years). Upper panels show the results from physical model 1, and lower panels show the results from physical model 2. The dashed lines show the density evolution and the upper x-axis shows the temperature variation with time. Observed abundances of molecules towards I16293B are displayed in panels by  solid-faint, colored lines with their widths representating 20\% uncertainties.}
\label{fig:models}
\end{figure*}

Figure \ref{fig:models} displays the chemical evolution of MC abundance (with respect to H$_2$) obtained from the two adopted physical models. The top panels corresponds to the physical Model~1 while the bottom ones to Model~2.
It is evident, based on Figure \ref{fig:models}, that MC forms during the warm up phase and thermally desorbs into the gas phase. 
 However, depending on the physical conditions of the models, the fractional abundances may vary. 
As introduced earlier, in our network we do not have gas-phase reactions which can directly produce MC.
Its gas phase abundance mainly comes from the thermal desorption of the species which forms in the solid phase through the radical-radical reaction of \chem{CH_3O} and \chem{NH_2CO}. 

\begin{figure*}
    \centering
    \includegraphics[width=8.0cm]{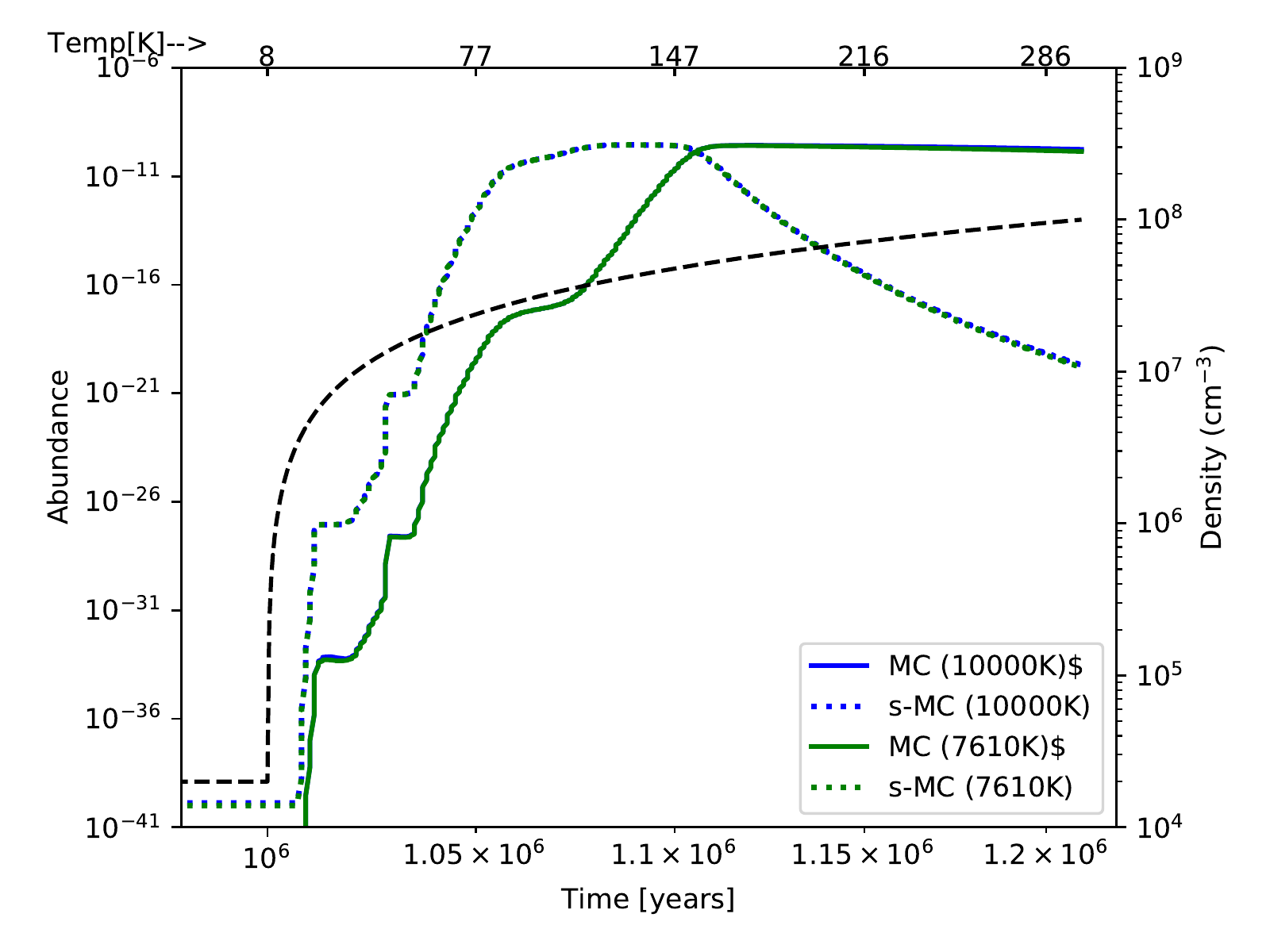}
    \includegraphics[width=8.0cm]{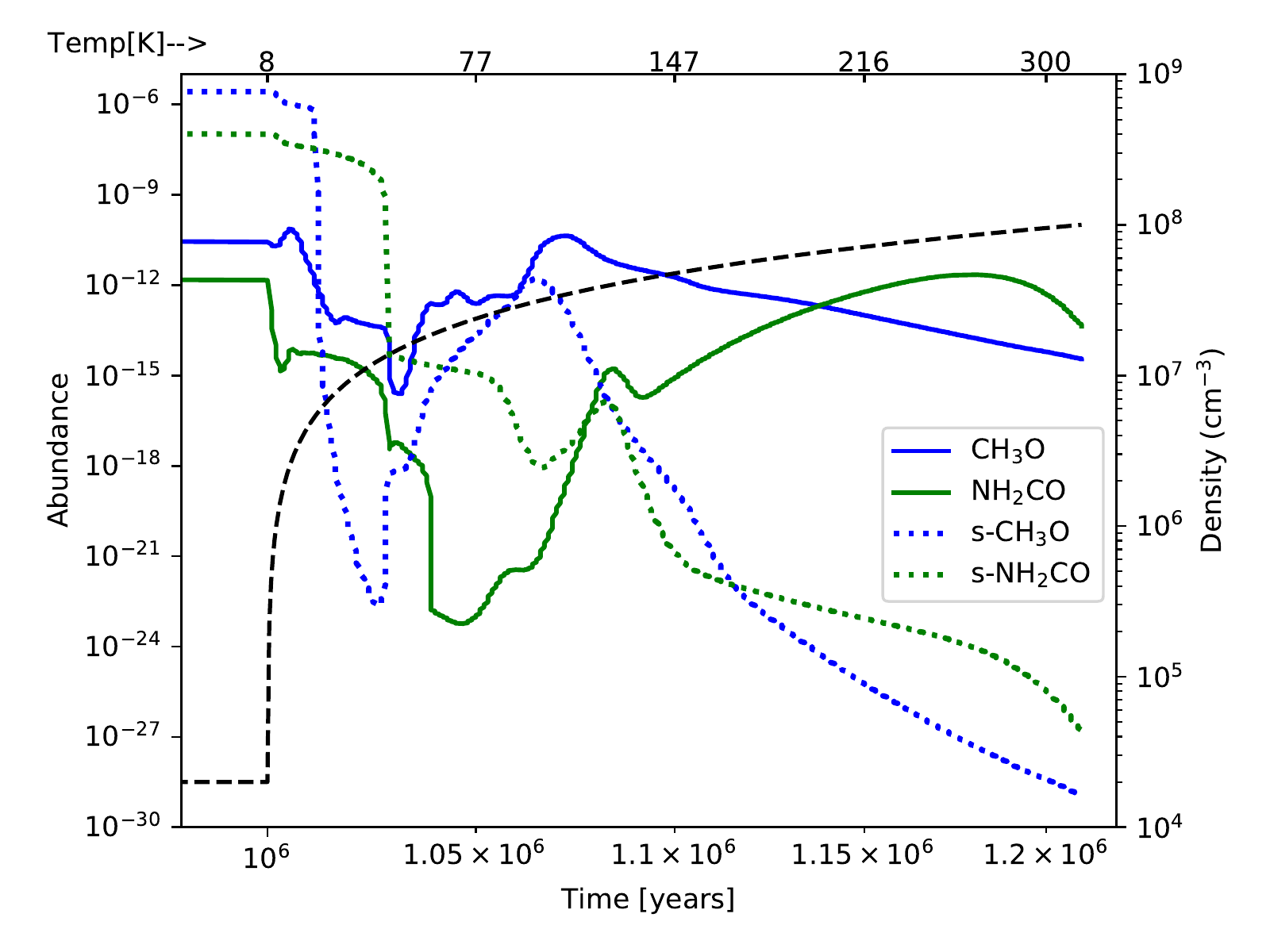}
    \caption{ Left: Evolution of MC is depicted for various values of binding energies ; Right: Evolution of major radicals, \chem{NH_2CO} and \chem{CH_3O} ( using Model 1 for both left and right panels}). Dotted lines show abundances in grain/dust surface and solid lines are for gas phase abundances. The dashed lines show the density evolution and the upper x-axis shows the temperature variation with time.  
    \label{fig:BE_var}
\end{figure*}

The binding energies greatly affects the mobility of radicals and consequently the formation of molecules on the grain surface.
Due to the high binding energies of \chem{CH_3O} and \chem{NH_2CO}, MC does not form effectively even in luke-warm ($\sim$30 K) conditions.
Furthermore, since we assume a high binding energy \citep[$\sim 10000$~K, considered similar to glycine, ][]{garrod13} for MC, it desorbs to the gas phase at high temperatures $>$100 K. 
As an example, for Model~1, MC starts major desorption at temperature $\sim150$ K while reaches it peak gas phase abundance at temperature $\sim 220$ K. 
These temperature values are highly dependent on the binding energy of MC. 
\citet{garrod13} considered the binding energy of MC to be 7610 K. 
To check the effect of binding energies, we consider two values 10000~K and 7610 K, and the results are shown in Figure~\ref{fig:BE_var}.  
For a binding energy of 7610 K, major desorption of MC occurs at a temperature $\sim 110$ K while it reaches its peak gas phase abundance at a temperature $\sim 160$ K. 
We can see that the evolution curve of MC shifts to the high-temperature side of the warm-up region for a higher value of binding energy.
So, the production of MC would be favored in hot-corino ($\sim$110~K) environments and below this temperature MC production is substantially low.

\begin{table}[]
    \centering
    \caption{  The fractional abundances of chemical species as obtained from chemical modeling.}
    \begin{tabular}{|l|l|l|}
    \hline
    Molecule & Model 1 & Model 2\\
    \hline
    \chem{CH_3OH}        &   1.1e-06  & 1.4e-06   \\
    \chem{CH_3CHO}       &   3.6e-08  &   1.4e-08 \\
    \chem{CH_3OCH_3}     &  1.3e-07   &  2.3e-07 \\
    \chem{CH_3NCO}     &  3.2e-10     &  1.2e-09 \\
    \chem{HNCO}       &    3.6e-09    &  1.8e-09 \\
    \chem{NH_2CHO}    &   2.7e-09     &  1.0e-09   \\
    \chem{H_2NCO_2CH_3}  &   1.7e-10  &  0.9e-10 \\
    \hline     
    \end{tabular}
    
    \label{tab:model_abun}
\end{table}

 To check the consistency of modeling compared to other well known COMs, we also show the chemical evolution of \chem{CH_3OH}, \chem{CH_3CHO} and \chem{CH_3OCH_3}.
These COMs are well abundant in hot-corino environments and have well-established pathways; therefore it is meaningful to compare the MC abundance with those of these COMs.
 \chem{CH_3OH} is the most abundant ($\sim 10^{-6}$) COMs in hot-corino environments.
Since the estimation of hydrogen density involves many uncertainties (mainly due to dust opacity), the abundances of COMs are often described comparing to \chem{CH_3OH}. 
Also, \chem{CH_3OCH_3} mainly forms on grain surfaces similar to the assumed grain-surface origin of MC, therefore it would be worthy to compare with MC. \chem{CH_3CHO} forms in both gas and grain phase. 
Additionally, interferometric observations of these molecules are available at comparable resolution toward the hot corinos that we discuss in this paper   \citep[e.g.,][and references therein]{sahu19,su19,droz19}.
The observational results help us to better constrain the chemical modeling.
In addition to these COMs, we consider \chem{NH_2CHO}, HNCO, \chem{CH_3NCO} which may be related to the formation of MC (see section~\ref{subsec:network}).

 Figure~\ref{fig:models} shows the chemical evolution of the molecules for two different sets of physical conditions.
For Model 1, we can see that the observed fractional column densities (w.r.t hydrogen) of molecules likely match the modeled abundances around a time $\sim 1.2 \times 10^6$, i.e $\sim 2 \times 10^5$ years of warm-up.  We estimated the best fit time focussing more on \chem{CH_3OH}, MC and \chem{NH_2CHO}. Though, for \chem{CH_3CHO} and \chem{CH_3OCH_3} best time may be near to $1.1 \times 10^6$ years but at this time model abundance of \chem{NH_2CHO} differs by orders of magnitude from observational value. So, we choose $1.2 \times 10^6$ years as the best-fit time for Model 1. The fractional abundances at this stage are noted in Table~\ref{tab:model_abun}.
 In the case of Model 2, the temperature quickly crosses over 100 K over a short time-scale ($\sim 10^4$ years). For Model 2, we noted down the abundances (Table~\ref{tab:model_abun}) after this short-time and within the final time of evolution.
The model abundances are quite similar in both models except for \chem{CH_3NCO}, which differ by an order of magnitude  from each other.
We note that the duration of the warm-up phase, in addition to the density profile, may severely affect the chemistry.
 The quicker and shorter warm-up in Model 2 affects the production of both \chem{CH_3NCO} and HNCO (formation pathways of which are closely linked) differently than in Model 1.
The results of modeling and observations are further compared and discussed in Section~\ref{discussion} .

\section{Observational Hints}

Earlier we \citep[Figure 2 of][]{sahu19} have found forest of spectral transitions towards the hot-corino object NGC1333 IRAS 4A2. 
From a subsequent spectral search, we speculated the possible signatures of methyl carbamate towards IRAS 4A2. 
To check whether the transitions of MC are present in other line rich sources, we searched ALMA archive for existing spectral observations in the range 349.8-351.6 GHz, the observed window towards NGC133 IRAS 4A2.
We found one of the well known hot-corino object, IRAS16293-2422 has been observed in the same frequency range (project id: ALMA\#2013.1.00278.S). 
The spectral emission towards the sources are strikingly similar and possible MC transitions (see Table~\ref{tab:transitions}) are present, too, in IRAS16293-2422 B. In this section, we briefly discuss the search of MC towards these hot corinos and also discuss the presence of relevant COMs.

 \begin{table*}
\centering
 \caption{ Methyl carbamate transitions }
\begin{tabular}{|c |c| c| c| c| } 
\hline
Species & Transition  (N Ka Kc v F) & Frequency (MHz)(Error) &  Eu (K) & \chem{A_{ij} }(s$^{-1}$)\\ [0.5ex] 
\hline
\hline
\chem{H_2NCO_2CH_3} & (20,15,5,3,20 - 19,14,6,3,19) & 349907.88 (0.0157)   & 322.69 & 9.22E-4  \\
\chem{H_2NCO_2CH_3} & (18,16,2,2,19 - 17,15,2,2,18) & 351056.46   (0.0098)    & 147.42 & 1.11E-3  \\
\chem{H_2NCO_2CH_3} & (18,16,3,1,18 - 17,15,3,1,17) &       351102.99 (0.0090)  &   147.38 & 1.11E-3 \\
\chem{H_2NCO_2CH_3} & (18,16,2,0,18 - 17,15,3,0,17) &   351162.83 (0.0092) &     147.42 & 1.11E-3  \\
\chem{H_2NCO_2CH_3} & (22,14,9,3,22 -21,13,8,3,21) & 351232.91 (0.0124) & 329.02  & 7.93E-4 \\ 
\hline

\multicolumn{5}{p{12cm}}{Note: This list contains only the main transitions that are discussed in the text,
transition-multiplicities are present around these frequencies.}
\label{tab:transitions}
\end{tabular}
\end{table*}

\begin{figure*}
    \centering
    \includegraphics[width=18cm]{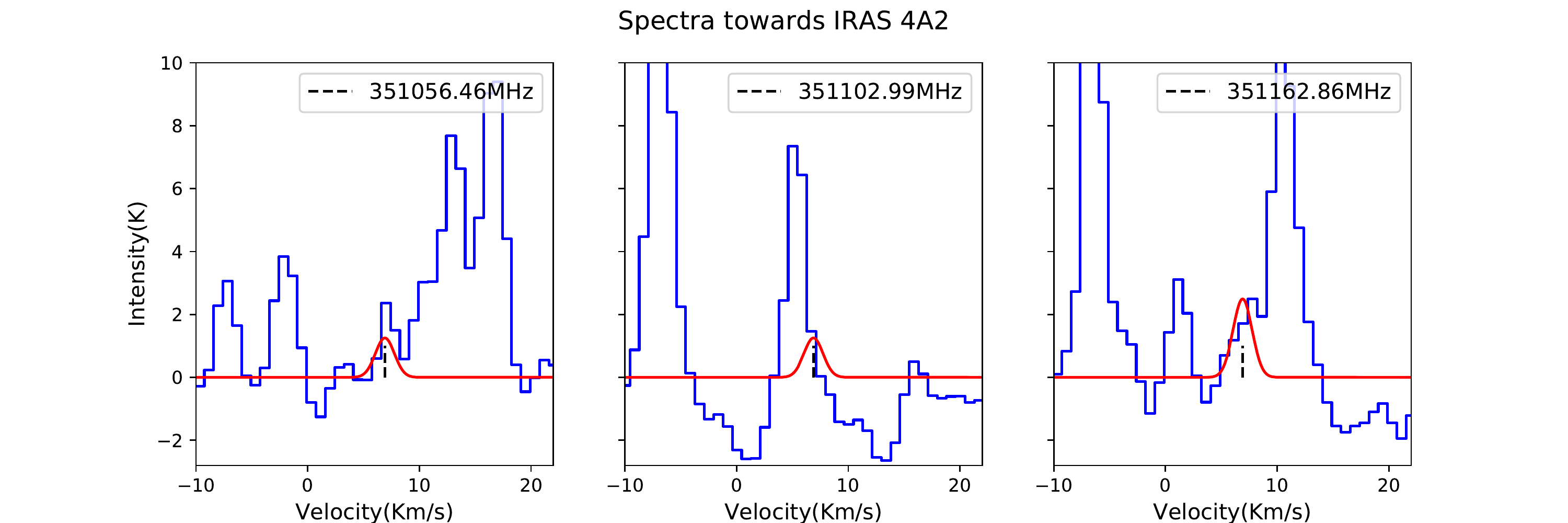}\\
    \vspace{0.5cm}
     \includegraphics[width=18cm]{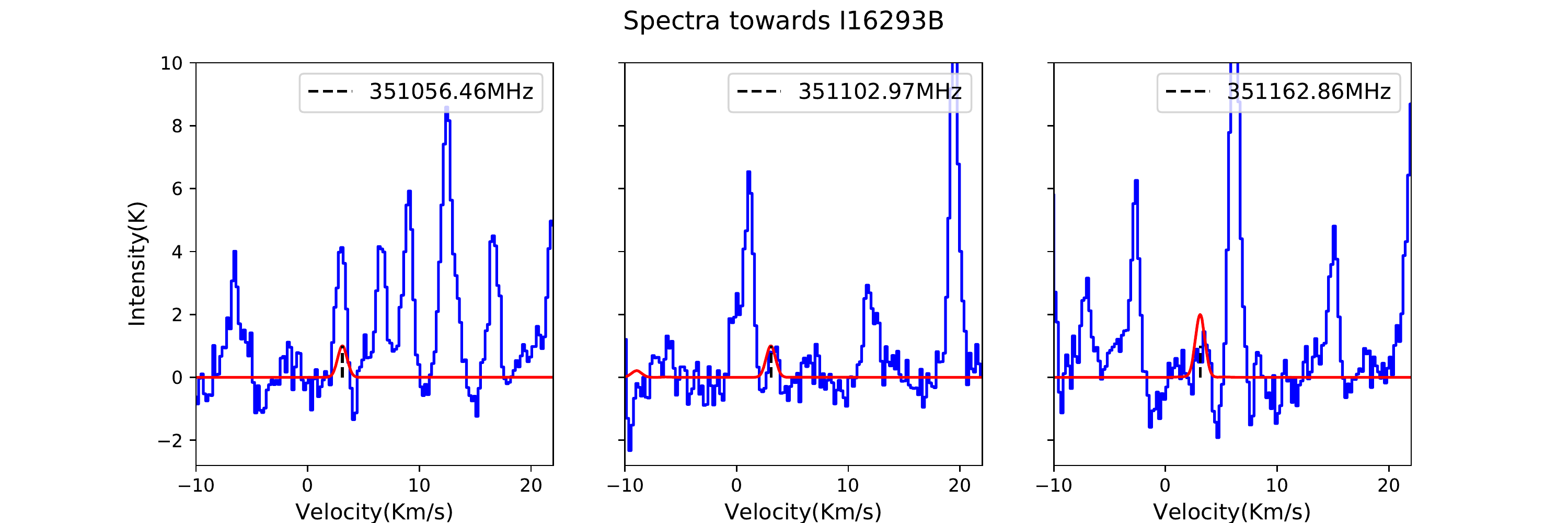}
    \caption{ The observed spectral transition of methyl carbamate tentatively detected towards IRAS 4A2 and I16293 are plotted in blue. The synthetic spectra estimated based on the LTE assumption are over-plotted in red. The dotted lines represent the systematic velocity, and the frequencies of transition are noted in the panels.}
    \label{fig:spec_lines}
\end{figure*}

\subsection{IRAS 4A2}
The NGC 1333 IRAS 4A (IRAS 4A hereafter) is one of the first known hot corino \citep{bott04} located in the Perseus molecular cloud at a distance of $\sim$ 293 pc \citep{ortiz18, zucker18}.
\citet{bott04} recognized IRAS 4A as a Class 0 source and reported the presence of  COMs, including \chem{HCOOCH_3}, HCOOH, and \chem{CH_3CN} based on IRAM -30m telescope observations at angular resolutions ranging from $10''$ to $24''$. 
As IRAS 4A is in fact a protobinary system consists of two protostellar sources IRAS 4A1 and IRAS 4A2 at a separation of 1.$''$8 \citep{looney00, reipurth02}, the  ``hot corino'' signatures from single dish observations by \citet{bott04} could not be attributed to the individual cores A1 and A2.
While later observations \citep{persson12} found that A2 harbors a hot corino, it is only recently that \citet{sahu19} discussed the possibility of A1 to host a hot-corino atmosphere.

In brief, the ALMA observations carried out in \citet{sahu19} and \citet{su19} deployed seven spectral windows, including one broadband window centered at 350.714 GHz whose bandwidth and spectral channel width are 1.875 GHz and 976 kHz ($\sim$ 0.84 \kms), respectively. 
The synthesized beam size of this spectral data cube is $\sim 0''.3 \times 0''.2$ (PA = $-6.45^{\circ}$), sufficient for resolving the two cores A1 and A2, and the rms noise level ($\sigma$) is 3 mJy~beam$^{-1}$ ($\sim 0.5$ K).
The dust-continuum emission towards IRAS 4A2 is predominately optically thin and from the 2D Gaussian fittings, the extent of the compact component of dust is found to be $\sim 0''.27$ \citep[see Table 1 of][] {sahu19}.  Assuming,  $\kappa_\nu$=0.0182 $\rm{cm^2g^{-1}}$ \citep{jorgensen16}, the average (from extended and compact emission) hydrogen column density is $\sim 2.5 \times 10^{25}$ \cms.

To analyze the spectra, we used CASSIS (developed by IRAP-UPS/CNRS- http://cassis.irap.omp.eu)
software as well as the JPL \citep{pick98} and CDMS \citep{muller01,muller05} databases. 
The spectroscopy data of MC that are available in JPL was studied by \citet{Ilyushin06}. 
The spectra towards IRAS 4A2 are extracted from a region similar to beam size around the peak position of dust-continuum. The line identifications are performed by assuming the systematic velocity to be 6.9 \kms \citep[] [and references therein]{sahu19}.
 In the appendix section we show the synthetic LTE spectra, in which at around 20 frequencies there are 106 MC transitions for the frequency range under consideration.
We find that for $T_{ex}=100$~K and 300~K, respectively, three and five emission features are at the level of 2$\sigma$ or at best 3$\sigma$, where $\sigma$ is rms noise of the observed spectra. There are multiple transitions around the frequencies of those emission features 
(see appendix, Table~\ref{tab:all_transitions}). Representative transitions around those emission features are  listed in Table~\ref{tab:transitions}. 
These transitions are useful for estimating the upper limit of the MC column densities in the observed spectra.
At $T_{ex}=100$~K, among the three emission profiles (see upper panel, Figure~\ref{fig:spec_lines}), one appears fully blended (351056.456 MHz) and the other emission features (351103.016 MHz, 351162.862 MHz) may be partially blended.
For $T_{ex}=300$~K, those three lines remain useful for estimating the upper limits. Additionally, two higher E$_u$ ($\sim 320$~K)  line appear in the spectra which are fully blended (see Figure in Appendix section B).
Column density estimations are made with those two temperature conditions. 
  
The Einstein coefficients (\chem{A_{ij}}) of all  those major transitions are found to be $> 7 \times 10^{-4}$ s$^{-1}$. 
The synthetic spectrum generated by CASSIS with a FWHM of $\sim 2.0$~\kms \citep[considering an average {Gaussian width $\sim$ 1~\kms}, as reported by][]{sahu19}, a column density of $2.5\times 10^{15}$~\cms, an excitation temperature of 100 K, and a source velocity of 6.9 \kms could reproduce the observed spectra. 
 The E$_u$ of these transitions are around $\sim$147 K, it is therefore difficult to estimate the rotational temperature. 
We choose another typical excitation temperature, 300 K, to estimate the column density. 
At this temperature, a column density $\sim 7\times 10^{15}$~\cms ~could reproduce the observed spectral features.
The transition around 351056 MHz is possibly blended with \chem{D_2C_2O}  and aGg$'$-glycol. 
 However, we could not quantitatively estimate the percentage of blending from these species as within the frequency range of the spectral window there are not enough transitions of those species to constrain their presence and abundances.
 The rest two transitions of MC are partially blended with intense transitions from other species, and the transition around 351162.862 MHz is estimated at a level $\sim$ 3$\sigma$. 
Considering these limitations, we only estimate the upper limit of MC from the observational data set.

It is relevant to mention the observed column densities of \chem{CH_3OH} and \chem{CH_3CHO} for comparison with methyl carbamate.
\citet{sahu19} found that \chem{CH_3OH} line emission may be optically thick and provide a lower limit for column density. So, to get the column density of \chem{CH_3OH}, we use the observe column density of  \chem{^{13}CH_3OH} \citep[see Table 3 of][] {sahu19} and use $^{12}$C/$^{13}$C ratio $\simeq 35$ \citep{jorgensen16}, the resultant column density is $1.12\times 10^{19}$~cm$^{-2}$. 
\chem{CH_3CHO} line emission is also found to be  optically thick, so to get the likely column density of it, we use \chem{CH_3OH}/\chem{CH_3CHO} ratio ($\sim 83$) towards I16293B  \citep[see listed column densities by][]{droz19}, and the estimated column density towards IRAS 4A2 found to be $1.35\times 10^{17}$~cm$^{-2}$; this value is consistent with the observed lower limit, $3.86\times 10^{16}$~cm$^{-2}$ towards IRAS 4A2.  Additionally, among other COMs, \chem{CH_3OCH_3} mainly produce in ISM on grain/ice surface,  found to be present  towards IRAS 4A2 \citep{lopez17}. Using an excitation condition from the earlier work (T$_{ex} \sim 130$~K), we found \chem{CH_3OCH_3} column density to be $\sim$ $9\times 10^{16}$~cm$^{-2}$. These observational values are helpful for comparing with the chemical model results.

\subsection{IRAS 16293}
IRAS16293-2422 (IRAS16293/I16293 hereafter) is a hot corino source similar to IRAS 4A and also a protobinary consists of IRAS 16293A and  IRAS 16293B.
The binary protostars are separated by 5.1$''$ \citep[620 AU; see] [ and references therein]{jorgensen16}. 
The source I16293B is chemically rich and the dust continuum is optically thick within a scale of 50 AU. 
We choose I16293B for our analysis as it is likely a face-on disk system and consequently the line widths are narrow. 
I16293A likely represents an edge-on system and its molecular line widths are comparatively broad.
\citet{jorgensen16} found that the lines (FWHM $\sim 1$~\kms) toward I16293B is five times narrower than I16293A, and this narrow line-width is helpful for detecting lines with less line confusion. 
I16293B therefore is an ideal target to search for MC.

The data towards the source were taken from ALMA archive and imaged with the highest possible spectral resolution using CASA pipeline analysis.
The whole data description can be found in \citet{jorgensen16} and we used only the 12m data, suitable to resolve the source without the missing-flux problem.  The synthesized beam size for the dust-continuum map is  $\sim 0''.63 \times 0''.39$ (Position angle = $-79.92^{\circ}$); this is similar to the  beam size  for spectral cube. 
 \citet{jorgensen16} argued that at its peak location, the dust continuum emission is optically thick and consequently the author chose a one-beam offset position from the continuum peak for extracting spectra. 
Assuming the dust emission at the offset position is optically thin, \chem{N_{H_2}} found to be  $\sim 1.2 \times 10^{25}$ \cms. Therefore, we choose a circular region similar to the observing 
beam ($0''.5$) around a position one offset away from continuum peak along the south west direction. Spectra from this position are expected to be least affected by the optical depth effect.
\citet{jorgensen16} reported a rms noise level of 7-10 mJy beam$^{-1}$ ($\le 0.6$K), where the channel width is 0.244 MHz ($\sim 0.2$ \kms).
With this educated estimation, we used the `statcont' algorithm \citep{sanchez18} to get the continuum subtracted emission. 
Line identification was done in same manner as described for IRAS 4A2, and we adopted a systematic velocity  of $\sim$ 3.1 \kms 
\citep{jorgensen11}.
We found that, for excitation temperatures of 100 K and 300 K, respectively, the column densities of $1.0\times 10^{15}$ cm$^{-2}$ and $2.3\times 10^{15}$ cm$^{-2}$ best match with the observed spectral features, assuming FWHM $\sim 1 $~\kms.
Additionally, the observed column densities for \chem{CH_3OH} ,\chem{CH_3CHO}  and \chem{CH_3OCH_3}  are $1.0\times 10^{19}$, $1.2\times10^{17}$~cm$^{-2}$, and $2.4\times10^{17}$ respectively \citep{jorgensen16,jorgensen18}.
Figure~\ref{fig:spec_lines} (lower panel) shows the tentative emission features of MC towards I16293B,  assuming an excitation temperature of 100 K. 

\begin{figure}
    \centering
   \includegraphics[width=9.0cm]{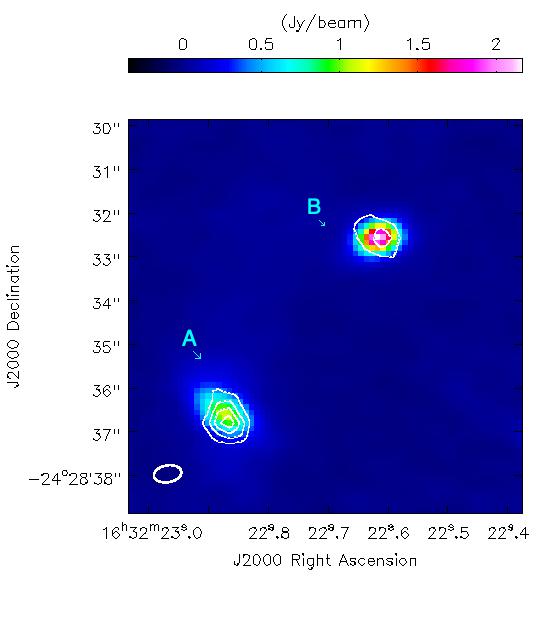}
    \caption{Integrated emission map (contours) of the 351103MHz MC transition towards I16293 overplotted with continuum emission map (color). We note that the integrated MC emission is localized around the hot corino I16293B. We note that toward I16293A, although the signal-to-noiase ratio is high, we cannot solely attributed the integrated emission to MC given the different Vlsr and line contamination. Contours are 5,10,15$\sigma$..., where $\sigma$=  6 mJy beam$^{-1}$ \kms.}
    \label{fig:image_IB}
\end{figure}

 We  note that for IRAS 4A2 all the spectral features of MC are either partially or fully blended due to low spectral resolutions.
On the other hand, for I16293B, the 351103~MHz \& 351162 MHz lines of MC seem to be resolved and non-blended. 
We choose the 351103~MHz line to check the emission map corresponding to this transitions. 
The integrated emission map related to the transition is shown in Figure~\ref{fig:image_IB}. 
Although the emission features presented in this paper are not sufficient yet for claiming a tentative detection of MC as there are only a limited number of transitions with their features being weak and blended.
These emission features, however, are sufficiently good for estimating an upper limit of MC.
 
\section{Discussions and conclusions}\label{discussion}

In this paper we present the astrochemical modeling of methyl carbamate and its search toward line-rich hot-corino sources IRAS 4A2 and I16293B. Under the physical conditions that are described in the text, we can closely reproduce the observed fractional abundance of molecular species.
Since obtaining the hydrogen column density from dust emission involves various uncertainties, we calculated the column density ratios of MC with respect to well known COMs, such as \chem{CH_3OH}, \chem{CH_3CHO},  \chem{CH_3OCH_3} assuming these COM species reside in the same volume.  For both sources, the hydrogen column densities are  $\sim 10^{25}$\cms, and methanol column densities are $\sim 10^{19}$\cms. Since the dust continuum is assumed to be optically thin, hydrogen column densities presented in the text are lower limits of hydrogen column densities. Therefore, the upper limit of methanol abundance (w.r.t. hydrogen) is of the order of $10^{-6}$. In the chemical modeling, for different physical conditions, we obtain methanol abundances of $\sim  10^{-6}$. Therefore, the results of the chemical models are suitable for describing the major abundant COM, methanol. We compare the observed fractional abundances of these COMs with MC in Figure~\ref{fig:barplot}. We can see that the observed and model abundances of MC comparing to those COMs match well within an order of magnitude. 
 This effectively gives us the idea of the MC abundance relative to those well abundant COMs. 
 
\begin{figure}
    \centering
    \includegraphics[width=8cm]{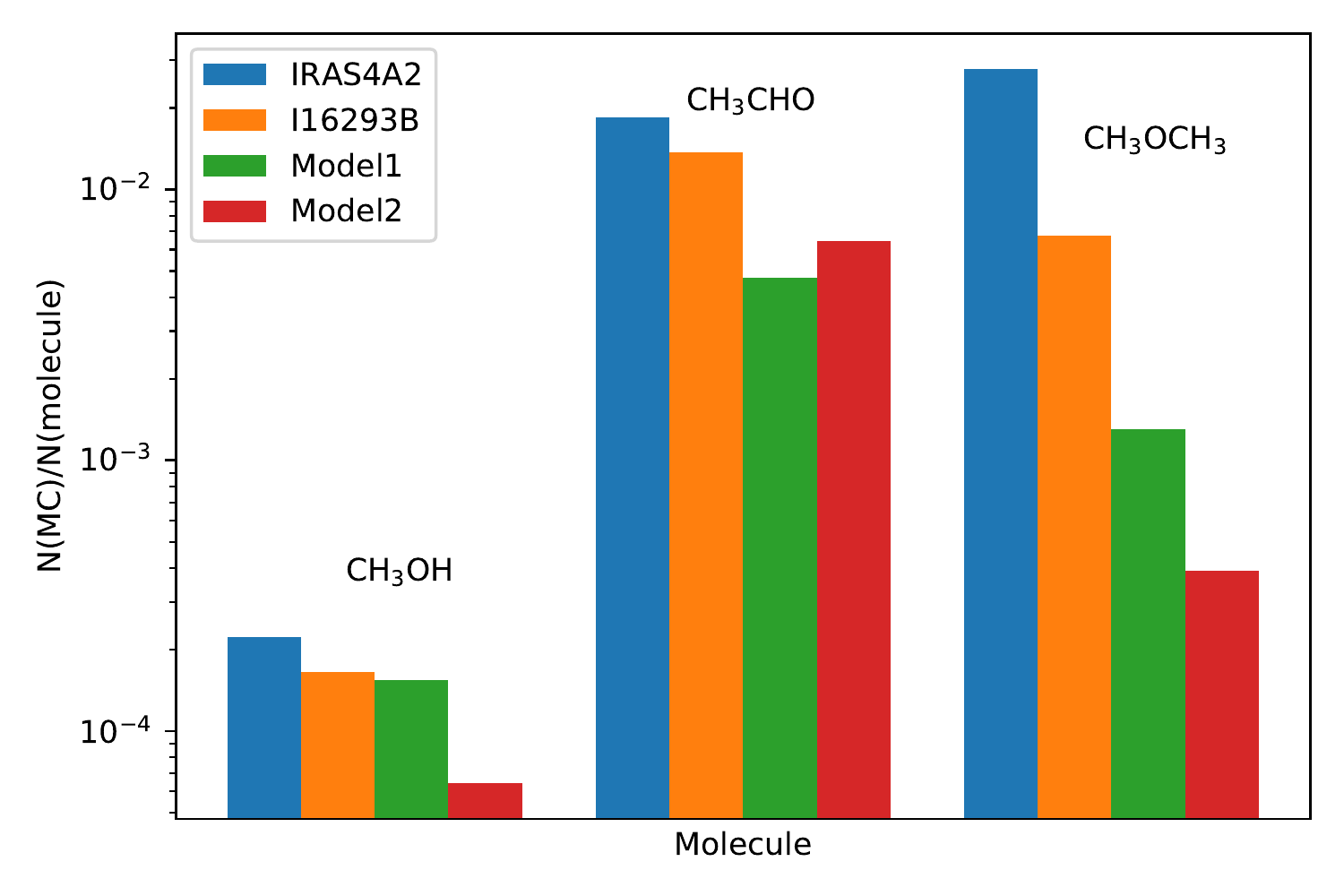}
    \includegraphics[width=8cm]{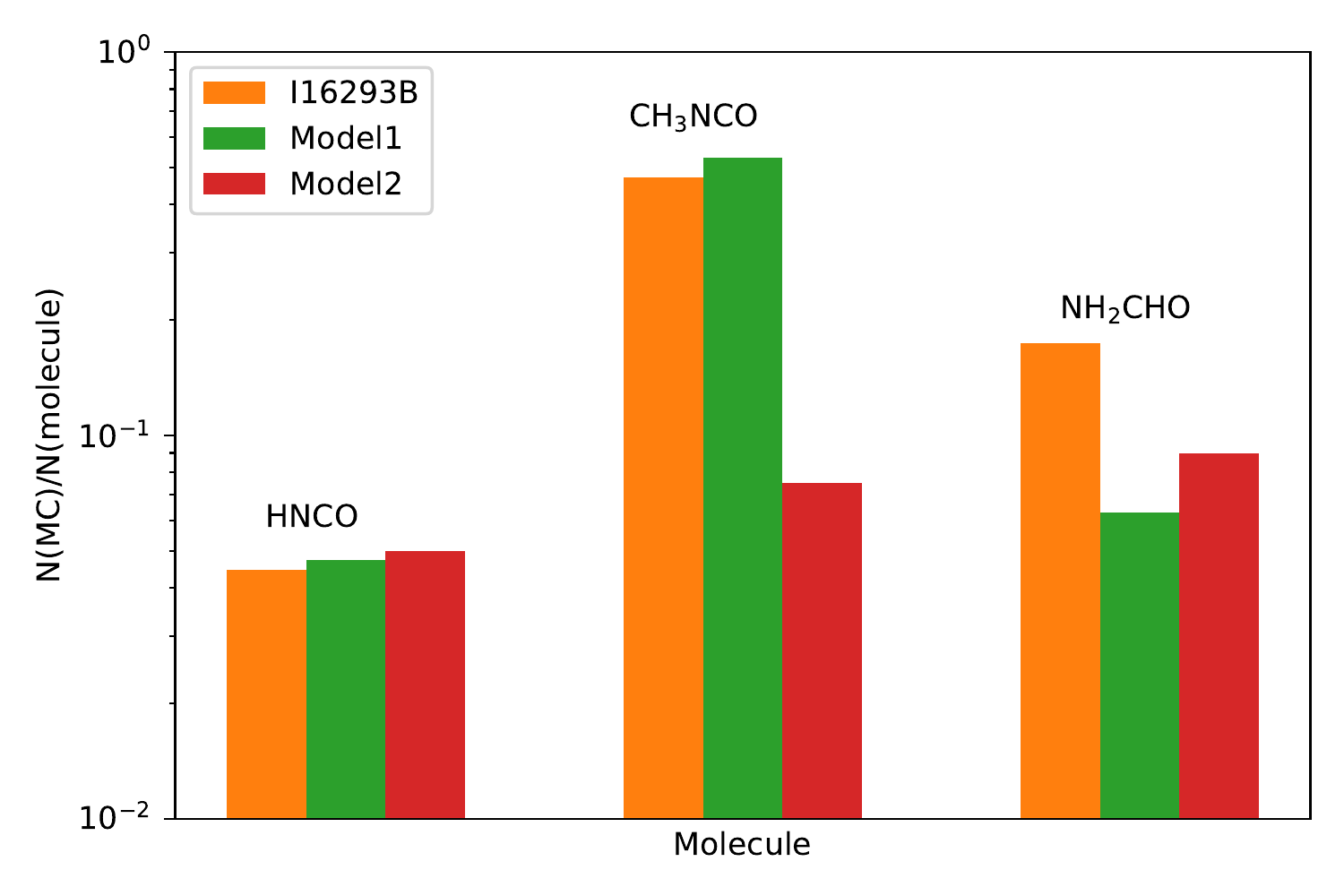}
    \caption{  Plots shows the relative abundance of methyl carbamate (MC) with respect to \chem{CH_3OH}, \chem{CH_3CHO} and \chem{CH_3OCH_3}  obtained from observations (I16293B, IRAS 4A2) and chemical modeling, Model 1  and Model 2. The observational values for MC are the average of estimated range of column densities.}
    \label{fig:barplot}
\end{figure} 

 Since one main constituent of MC is \chem{NH_2CO} radical, we consider species- \chem{CH_3NCO}, HNCO, \chem{NH_2CHO} which are closely linked with \chem{NH_2CO} and therefore MC. 
 Towards I16293B, their abundances have been measured at a comparable resolution  \citep[see][and references therein]{droz19}. 
For IRAS 4A2, some of these species have been observed at much coarser angular resolutions  \citep[e.g.,][and references therein]{taquet15}, and there may be large uncertainties in column density estimation due to  beam dilution factor.
Therefore, comparison between the model and observed abundances are shown only for I16293B (Figure~\ref{fig:barplot}).

 We can see that the modeled and  observed fractional abundances of MC with respective to \chem{CH_3OH} and \chem{CH_3CHO} closely match. 
 These molecules are major abundant molecules in hot corinos, so the abundance ratio of these molecules to MC can be helpful to gauge the MC abundances in hot-corino environments.
\chem{CH_3OCH_3} is overproduced in our chemical modeling, resulting in a order of magnitude difference between observed and modeled abundance. 
As the radical \chem{CH_3O}, a major constituent of MC, is mainly produced with reactions related to \chem{CH_3OH}, so \chem{CH_3OCH_3} may not be a good species for estimating the MC abundance. 
 On the other hand, modeled abundances ratios of MC and species like HNCO, \chem{CH_3NCO}, \chem{NH_2CHO} are mostly in agreement with the observed results except for \chem{CH_3NCO} estimation of Model 2. 
Due to very short and quick warm-up time scale of Model 2, this kind of differences may arise. 
However, the results suggest that  HNCO, \chem{CH_3NCO}, \chem{NH_2CHO} maybe helpful for estimating MC abundances.
 Significance of the modeling results is that the upper limit of observed column densities ($\sim 10^{15}$ cm$^{-2}$) of MC (an glycine isomer)  roughly consistent with  the modeled abundances. For the first time, we performed the observational search of MC and chemical modeling. Results of chemical modelling suggest that species like \chem{CH_3OH}, HNCO, \chem{CH_3NCO}, \chem{NH_2CHO} may helpful for estimating possible abundance of MC, using this estimations MC can be searched toward hot corinos. 

MC can be searched using modern telescope facility like ALMA and future telescopes like ngVLA.
To avoid line confusion problem, MC can be searched using high sensitive ALMA spectral line observations in lower frequencies ( 84 -116 GHz, Band 3).  However, it should be noted that for larger molecules like MC, the transitions are very weak in the range of ALMA's spectral coverage. Rotational transitions of lighter molecules mostly fall within  the mm-submm spectral range, so spectral confusion limit may reach in short time ($\sim$1 hour) and it may make identification of MC difficult. Since, rotational transition of comparatively smaller molecules (e.g.,\chem{CH_3OH}) do not fall in centimeter wavelength range, this wavelength regime is comparatively clearer and therefore reduces the level of line-confusion. Therefore, future astronomical telescope like ngVLA 
 may be helpful for detecting the glycine isomer, MC. Using the MC column-density  upper limit, we find that the strongest transition around 43 GHz have intensity $\sim 0.3$~mJy. Assuming, beam size fill the source, a sensitivity of $\sim 0.05$~mJy can be achieved by ngVLA approximately in 10 hours of integration time \citep{mcguire18}. Therefore, it is possible to detect MC using ngVLA
 toward the hot corinos. Detection of MC in the ISM may shed light on the long standing problem to understand glycine/prebiotic chemistry in the ISM.

 {\bf Acknowledgment:}
 We thank the reviewers for their constructive comments and suggestions. D.S. acknowledges ASIAA post doctoral research fund, also thanks Dr. Bhalamurugan Sivaraman for helpful discussions. S.Y.L. acknowledges the support by the Minister of Science and Technology of Taiwan (MOST 107-2119-M-001-041). A.D. would like to acknowledge ISRO respond project (Grant
No. ISRO/RES/2/402/16-17) for partial financial support and Grant-In-Aid from the Higher Education Department of the Government of West Bengal. 
This paper makes use of the following ALMA data:  ADS/JAO.ALMA\#2015.1.00147.S \& ALMA\#2013.1.00278.S. ALMA is a partnership of ESO (representing its member states), NSF (USA) and NINS (Japan), together with NRC (Canada) and MoST and ASIAA (Taiwan) and KASI (Republic of Korea), in cooperation with the Republic of Chile. The Joint ALMA Observatory is operated by ESO, AUI/NRAO and NAOJ. 
 

\appendix

\renewcommand\thefigure{\thesection.\arabic{figure}} 
\setcounter{table}{0}
\renewcommand\thetable{\thesection.\arabic{table}} 

\section{Synthetic LTE spectra to search for major spectral transitions of MC}
Figure~\ref{fig:app1} shows the synthetic LTE spectra (using CASSIS software) of MC for column densities of $1.0\times10^{15}$ cm$^{-2}$ and $2.3\times10^{15}$  cm$^{-2}$,  respectively for excitation temperatures of 100~K and 300~K towards I16293B; source size was assumed to be 0.$''$5. These figures give an idea about the major emission signatures of MC to look for. For T$_{ex}$=100~K and 300~K, respectively, there are three and five major transitions at a label of 2$\sigma$ (rms) or higher. These transitions are listed in Table~\ref{tab:transitions}. Based on these transitions, the upper limit of MC column densities are estimated from the observed spectra towards the hot corinos. 

\begin{figure}[ht]
   \setcounter{figure}{0}
    \centering
    \includegraphics[width=17cm]{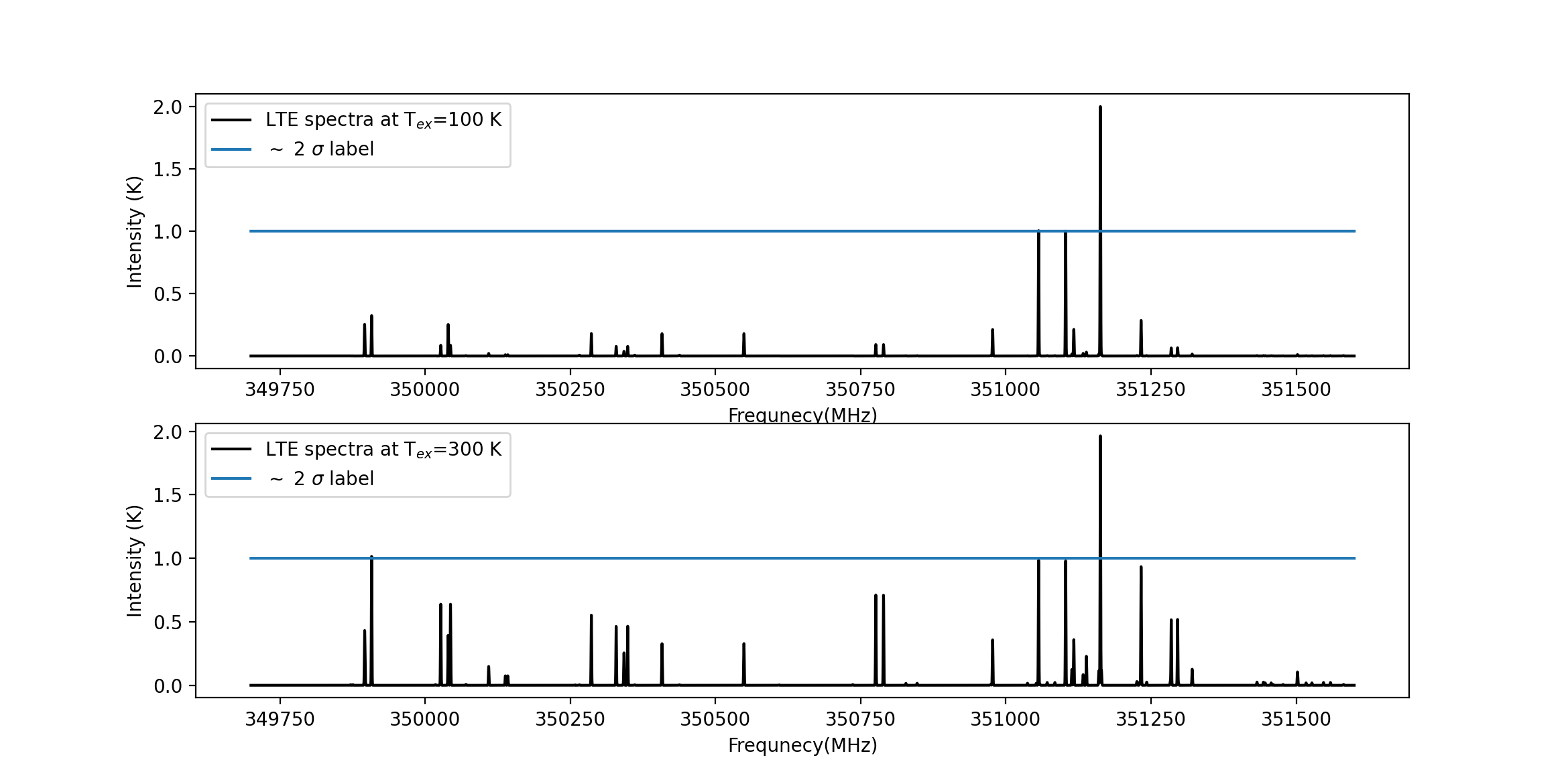}
    \caption{A typical LTE synthetic spectra is displayed considering excitation temperature 100K and 300K. Straight lines shows 2$\sigma$ level of the observed spectra towards IRASI16293B. }
    \label{fig:app1}
\end{figure}

\section{LTE spectra for T$_{ex}$=300 K }
Figure~\ref{fig:app2} and \ref{fig:app3} show the LTE synthetic spectra of MC towards IRAS 4A2 and I16293B respectively, for an excitation temperature of 300~K. We searched for all MC lines having E$_u <$ 500~K. As the emission features of MC are very weak and blended, the rotational temperature can not be estimated from the observed spectra. Therefore, we assumed two typical temperatures, 100~K and 300~K, for estimating the upper limit of MC column densities. The LTE spectra for 100~K are already shown in figure~\ref{fig:spec_lines}. We can see that for T$_{ex}$=300 K  few high-temperature lines appear in the spectra. However, these high E$_u$ lines (349907.88 MHz, 351232.91 MHz) are fully blended but it helps us estimate the range of column density upper limits. 

\begin{figure}[ht]
    \centering
     \setcounter{figure}{0}
    \includegraphics[width=18cm]{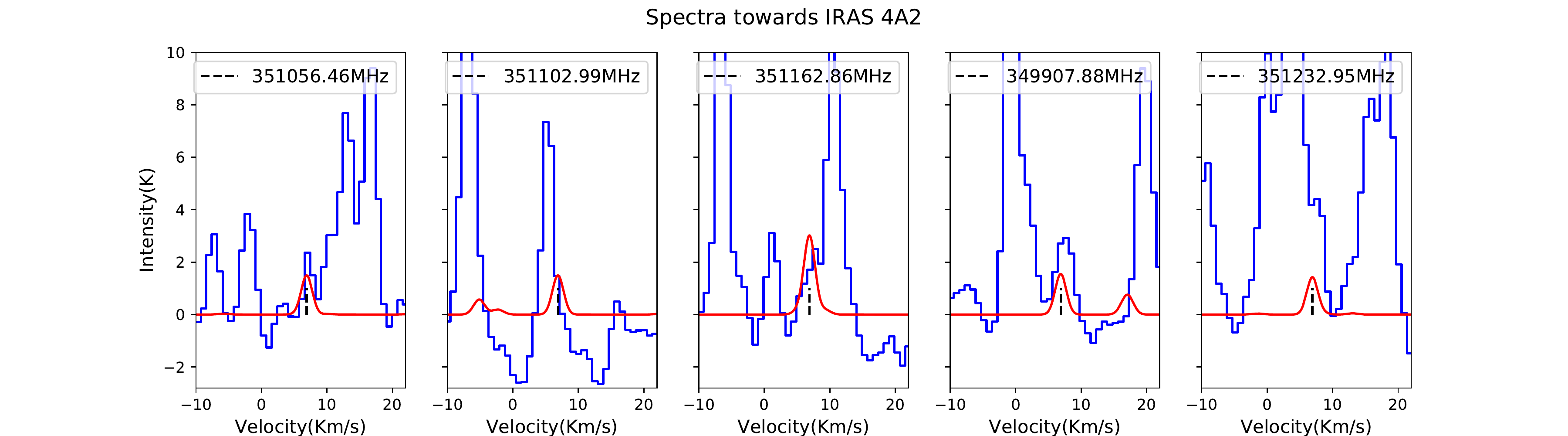}
    \caption{LTE synthetic spectra (red) for an excitation temperature of 300~K overplotted on observed spectra (blue) towards IRAS 4A2. }
    \label{fig:app2}
\end{figure}

\begin{figure}[ht]
    \centering
     \includegraphics[width=18cm]{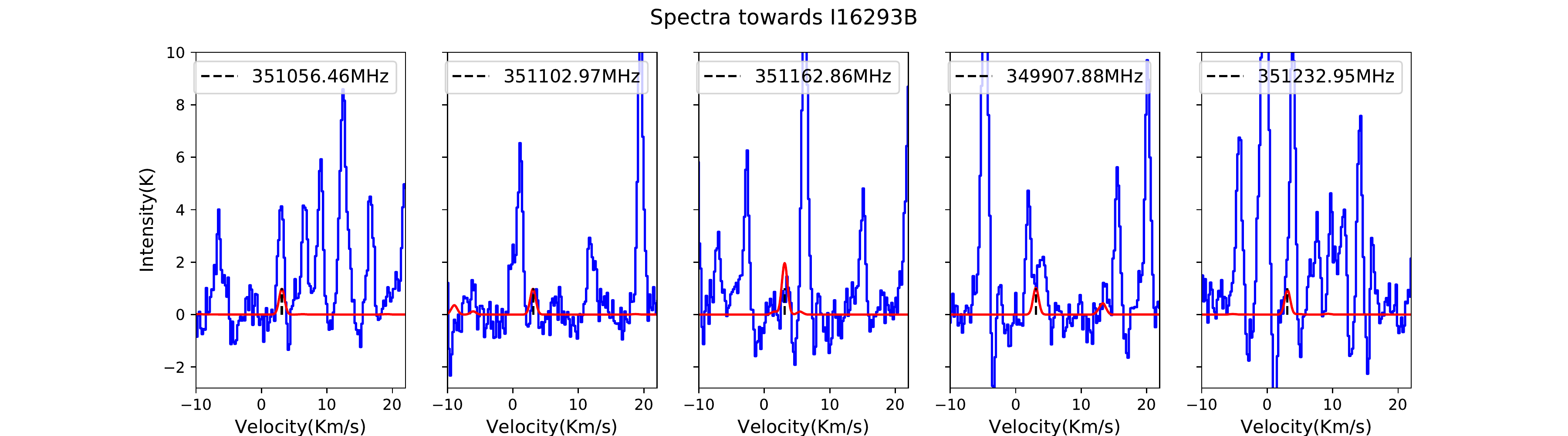}
    \caption{Similar to Figure~\ref{fig:app2}, spectra towards IRAS I16293B.}
    \label{fig:app3}
\end{figure}

\section{Full list of MC transitions}
The Table~\ref{tab:transitions} in the text shows the list of representative transitions. Multiple transitions are present around those frequencies. Table~\ref{tab:all_transitions} list multiple transitions around the frequencies. 
\begin{table*}
\centering
 \caption{ Methyl carbamate transitions }
   \begin{tabular}{|c |c| c| c| c|} 
\hline
Species & Transition  (N Ka Kc v F*) & Frequency (MHz) (Error) &  Eu (K) & \chem{A_{ij} }(s$^{-1}$) \\ 
\hline
\hline
\chem{H_2NCO_2CH_3} & (20,15,5,3,20 -19,14,6,3,19) &  349907.88 (0.0157) &    322.69 & 9.22E-4 \\
\chem{H_2NCO_2CH_3} & (20,15,6,3,20- 19,14,5,3,19) &   349907.88 (0.0157) &     322.69 & 9.22E-4 \\
 \chem{H_2NCO_2CH_3} & (20,15,5,3,21 - 19,14,6,3,20) &  349907.92   (0.0157) &      322.69 & 9.25E-4 \\
\chem{H_2NCO_2CH_3} & (20,15,6,3,21 - 19,14,5,3,20) &   349907.92 (0.0157) &    322.69 & 9.25E-4 \\
\chem{H_2NCO_2CH_3} & (20,15,5,3,19 - 19,14,6,3,18) &        349907.92 (0.0157) &    322.69 &9.22E-4 \\
 \chem{H_2NCO_2CH_3} & (20,15,6,3,19 -19,14,5,3,18)  &     349907.92 (0.0157) &    322.69 & 9.22E-4 \\
 \hline
\chem{H_2NCO_2CH_3} & (18,16,2,2,18 - 17,15,2,2,17) & 351056.43   (0.0098)  & 147.42 & 3.29E-3 \\
\chem{H_2NCO_2CH_3} & (18,16,2,2,17 - 17,15,2,2,16)  &  351056.46  (0.0098)&     147.42 & 3.11E-3 \\
\chem{H_2NCO_2CH_3} & (18,16,2,2,19 - 17,15,2,2,18) & 351056.46  (0.0098)    & 147.42 & 1.11E-3  \\
\hline
\chem{H_2NCO_2CH_3} & (18,16,3,1,18 - 17,15,3,1,17) &       351102.99  (0.0090) &   147.38 & 1.11E-3 \\
\chem{H_2NCO_2CH_3} & (18,16,3,1,17 - 17,15,3,1,16) &       351103.01 (0.0090)& 147.38 & 3.12E-3 \\
\chem{H_2NCO_2CH_3} & (18,16,3,1,19 - 17,15,3,1,18) &  351103.02  (0.0090)   & 147.38 & 3.48E-3  \\
\hline
\chem{H_2NCO_2CH_3} & (18,16,2,0,18 - 17,15,3,0,17) &   351162.83 (0.0092)&     147.42 & 1.11E-3   \\
\chem{H_2NCO_2CH_3} & (18,16,3,0,18 - 17,15,2,0,17) &        351162.83 (0.0092) &     147.42 & 3.29E-3 \\
\chem{H_2NCO_2CH_3} & (18,16,2,0,17 - 17,15,3,0,16)  &      351162.86 (0.0092) &    147.42 & 3.12E-3 \\
\chem{H_2NCO_2CH_3} & (18,16,3,0,17 - 17,15,2,0,16)  &   351162.86  (0.0092)   & 147.42 & 3.12E-3  \\
\chem{H_2NCO_2CH_3}& (18,16,2,0,19 - 17,15,3,0,18) &    351162.86 (0.0092)&     147.42 &  3.48E-3   \\
\chem{H_2NCO_2CH_3} & (18,16,3,0,19 - 17,15,2,0,18) &    351162.86 (0.0092) &     147.42 & 3.48E-3  \\
\hline
\chem{H_2NCO_2CH_3} & (22,14,9,3,22 -21,13,8,3,21) & 351232.91 (0.0124) & 329.02  & 7.93E-4 \\
\chem{H_2NCO_2CH_3} & (22,14,8,3,22 - 21,13,9,3,21)  &      351232.91 (0.0124) & 329.02 & 7.93E-4 \\
\chem{H_2NCO_2CH_3}  & (22,14,9,3,23 -21,13,8,3,22) & 351232.95 (0.0124) &   329.02 & 7.95E-4 \\
\chem{H_2NCO_2CH_3} & (22,14,8,3,23 -21,13,9,3,22) & 351232.95 (0.0124) &  329.02 & 7.95E-4 \\
\chem{H_2NCO_2CH_3} & (22,14,8,3,21 -21,13,9,3,20)  &  351232.95 (0.0124) &   329.02  & 7.93E-4 \\
\chem{H_2NCO_2CH_3} & (22,14,9,3,21 - 21,13,8,3,20)  & 351232.95 (0.0124) & 329.02 & 7.93E-4 \\

\hline
\multicolumn{5}{p{12cm}}{*The quantum numbers (QN) are mentioned as N Ka Kc v F .  The `F' QN can also denote an A or E state as in the case for MC.}
     \end{tabular}
     \label{tab:all_transitions}
        \end{table*}

\end{document}